\documentclass[aps,preprint,superscriptaddress,12pt,longbibliography]{revtex4-1}

\usepackage{amsmath}
\usepackage{graphicx}
\usepackage{subcaption}
\usepackage[labelfont=bf]{caption}
\usepackage[colorlinks]{hyperref}
\usepackage{float}

\begin{document}

\title{Single Chain Expulsion from Diblock Copolymer Micelles with Dense Corona}

\author{Shuang Yuan}
\affiliation{South China Advanced Institute for Soft Matter Science and Technology, School of Emergent Soft Matter, South China University of Technology, Guangzhou 510640, China}

\author{Jiajia Zhou}
\email[]{zhouj2@scut.edu.cn}
\affiliation{South China Advanced Institute for Soft Matter Science and Technology, School of Emergent Soft Matter, South China University of Technology, Guangzhou 510640, China}
\affiliation{Guangdong Provincial Key Laboratory of Functional and Intelligent Hybrid Materials and Devices, South China University of Technology, Guangzhou 510640, China}
\affiliation{State Key Laboratory of Pulp and Paper Engineering, South China University of Technology, Guangzhou 510640, China}

\begin{abstract}
We use self-consistent field theory to investigate the free energy landscape for single-chain expulsion from a diblock copolymer micelle with a dense corona. 
Using the distance from the micelle center-of-mass to the hydrophilic-hydrophobic junction of the chain as the reaction coordinate, we compute the free energy landscape for chain exchange. 
Our results show that the expulsion free energy barrier scales linearly with both the hydrophobic block length and the solvent selectivity, consistent with recent experiments.
To accurately resolve chain conformation, we introduce a second reaction coordinate: the distance between the junction and the free end of the hydrophobic block, and construct a two-dimensional free energy surface. 
Using the string method to identify the minimum energy path, we find that all pathways converge to a nearly degenerate reaction channel, irrespective of the initial path. 
Within this channel, the end-to-end distance of the hydrophobic block exhibits a broad distribution, yet the corresponding expulsion barriers remain nearly indistinguishable.
%Within the continuum SCFT framework, this work recovers the linear scaling of the chain expulsion barrier and the stretched transition state conformation observed in dissipative particle dynamics simulations.
%Furthermore, it reveals the degeneracy of transition state conformations and shows that the reaction channel broadens and shifts outward with increasing hydrophobic block length.
%The consistency between our theoretical model and particle-based simulations provides strong support for the hyperstretching mechanism under these conditions. 
Together, these findings establish a continuum-level theoretical foundation for understanding the hyperstretching mechanism and the transition state ensemble in micellar chain exchange.
\end{abstract}

\maketitle

%%%%%%%%%%%%%%%%%%%%%%%%%%%%%%%%%%
\section{Introduction}
\label{sec:intro}

Diblock copolymers self-assemble in selective solvents to form micelles of various nanostructures, such as spherical micelles, wormlike micelles, or vesicles \cite{Fontell1981, Lewis2018}. 
Their broad applications in drug delivery \cite{Jeffrey2003}, nanoreactors \cite{Cotanda2012}, and nanolithography \cite{Lohmueller2011} have made them a focus of research in soft matter science.
In solution, micelles are not static but exist in a state of continuous evolution. 
This dynamics is characterized by a constant exchange of polymer chains between the micelles and the surrounding medium: free chains insert into micelles while constituent chains exit to become free chains. 
Consequently, micelles exist in a dynamic equilibrium, undergoing continual adjustments in their aggregation number \cite{Daza2017}. 
The exchange of individual chains between micelles is governed by a two-step mechanism of expulsion from the core and subsequent insertion. 
The expulsion step is widely regarded as the rate-limiting step \cite{Choi2010}.
The rational design and precise regulation of functional micellar materials require a detailed understanding of chain exchange mechanism. 
This necessitates elucidating the relationship between the exchange dynamics and molecular parameters (e.g., block length, interaction parameters), as well as characterizing the structure and energy of the key transition states.

\begin{figure}[htbp]
  \includegraphics[width=0.5\columnwidth]{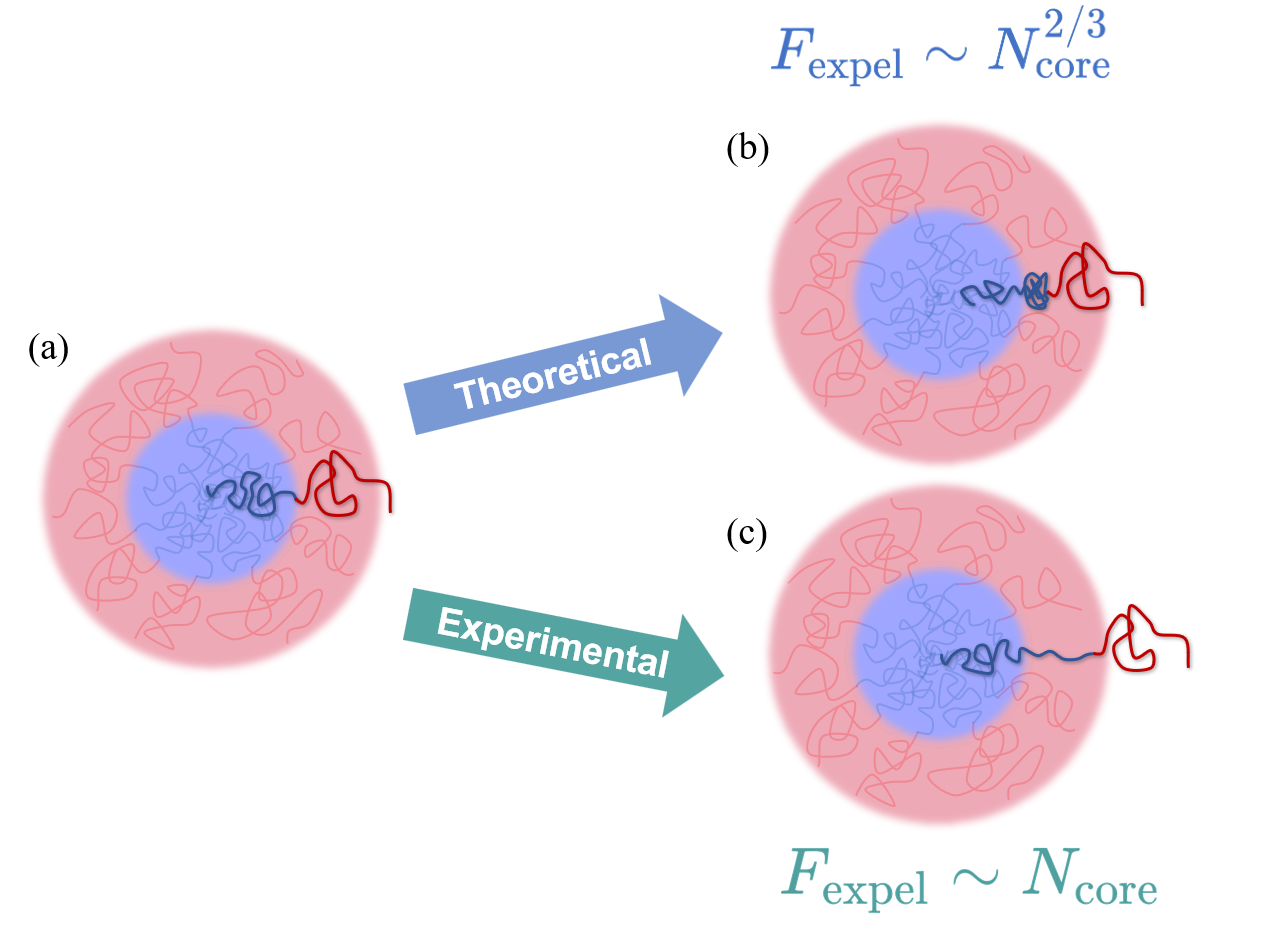}
  \caption{
    Discrepancy between theoretical and experimental scaling of the free energy barrier for chain exchange in micelles.
    (a) The expelled chain located in the micelle.  
    (b) The theoretically predicted budding conformation at the energy minimum.  
    (c) The stretched conformation observed in experiments.
    } 
  \label{fig1} 
\end{figure}

From a theoretical perspective, the seminal scaling model introduced by Halperin and Alexander (HA) \cite{Halperin1989} remains the primary framework for the kinetics of chain exchange.
HA model assumes that the core block collapses into a globule when entering the corona during chain expulsion, shown schematically in Figure \ref{fig1}(b).
Assuming the core block has $N_{\mathrm{core}}$ number of Kuhn segments, the collapsed state has a characteristic size scaling as $N_{\mathrm{core}}^{1/3}$. 
The free energy barrier arises from the unfavorable interactions, proportional to the surface tension $\gamma$ and the globule's surface area $N_{\mathrm{core}}^{2/3}$. 
This leads to the HA scaling of the free energy barrier as $\gamma N_{\mathrm{core}}^{2/3}$.
However, recent time-resolved small-angle neutron scattering (TR-SANS) experiments \cite{WangEn2020, WangEn2018, Choi2011, LuJie2016, LuJie2015, Lund2011, ZhaoDan2018, Zinn2012, LuJ2012, Peters2016, Kim2024} have revealed that the free energy barrier has a linear dependence on $N_{\mathrm{core}}$. 
The linear scaling corresponds to an extended state of the core block, shown in Figure \ref{fig1}(c).
These experimental findings implied a different microscopic escape mechanism that differs from the hypothesis of the HA model.

To elucidate the microscopic mechanism of chain expulsion, Seeger \emph{et al.} \cite{Seeger2021, Seeger2022} employed dissipative particle dynamics (DPD) \cite{Daza2017, Prhashanna2016, LiZhenlong2011, Prhashanna2020, Peters2017} combined with umbrella sampling \cite{Torrie1977, Giacomo2013} to investigate the process. 
Their study revealed that the core chain in the transition state does not fully collapse but rather adopts a ``hyperstretching'' conformation. 
In this case, the surface area scales as $N_{\mathrm{core}}$, leading to a free energy barrier that scales linearly with $N_{\mathrm{core}}$, consistent with experimental observations.

Recent study by Varner \emph{et al.} \cite{Varner2025} has further revealed the complexity of transition state conformations.
Through the analysis of two-dimensional free energy landscapes, two competing mechanisms during chain expulsion have been revealed: the collapse pathway corresponding to the HA mechanism and the hyperstretching pathway.
Under low-density conditions, the minimum free energy path corresponds to the HA mechanism of collapsed globule, exhibiting an $N^{2/3}_\mathrm{core}$ scaling of the barrier.  
In dense corona or high-density conditions, chains tend to escape through the hyperstretching mechanism, showing linear scaling of the free energy barrier.  
This demonstrates that both the transition state conformation and barrier scaling depend on the system's environmental density, corona chain crowding, and core chain collapse propensity.
These simulations provide valuable molecular-level insights into chain expulsion. 
%However, the standard DPD models \cite{Groot1997}\cite{Espanol1995} may not fully capture the conformational details of the stretching-collapse transition in continuous flexible chain systems at the transition state.  

Self-consistent field theory (SCFT) \cite{Matsen2002, QiangYC2020} is another choice in computing chain exchange energy landscapes. 
As a continuum field-theoretic approach, SCFT efficiently obtains chain conformational distributions and free energies by solving a modified diffusion equations.
SCFT has been successfully applied to systems with high degrees of polymerization and complex phase behaviors 
\cite{LiuMj2016, LiQY2021, XieN2014, Morgan2019}.
A SCFT framework was previously employed by Mysona \cite{Mysona2019, Mysona2019a, Mysona2019b, Mysona2020}, providing early insights into mapping the free energy for chain exchange into micelles.
Compared to particle-based simulations, the SCFT method offers easy access to free energy profiles. 
It avoids the statistical uncertainties that are inherent in sampling-based free energy methods.

This study aims to calculate the free-energy barrier of single chain expulsion for micelles with dense corona.
It is important to acknowledge that our current framework (detailed in Computational Method section), is intrinsically unable to reproduce the coil-to-globule collapse of the hydrophobic block in the solvent. 
As a result, this method is not positioned to prove or disprove the HA mechanism, as that scenario requires a transition to a compact, collapsed state that is not accessible within the chosen theoretical framework. 
However, our model is physically appropriate for describing micelles with a thick and dense corona, where the high-density environment acts like a polymer melt and suppresses complete globular collapse. 
The primary purpose of this study is to utilize this continuum framework to systematically explore the broad parameter space and determine the scaling behavior of the energy barrier for different molecular parameters, such as block length and solvent selectivity.

Our paper is organized as follows. 
In Section \ref{sec:method}, we briefly introduce the SCFT model and the methods used to compute the free energy profile for chain exchange. 
%In our calculations, we adopt a frozen-field approximation. 
%We compute the equilibrium chemical potential field of the micelle using a full SCFT calculation. 
%The field is kept fixed while we compute the free energy of a single chain at different reaction coordinates. 
%The approximation is justified because single chain expulsion occurs on a much faster time scale than the structural relaxation of the micelle. 
%It also avoids the need for statistical sampling, allowing us to construct the free energy landscape efficiently.
In Section \ref{sec:result}, we first discuss the thermodynamic stability of the micelles. 
We then select the distance from the micelle center-of-mass to the diblock junction as the initial reaction coordinate. 
Along this coordinate, we examine the dependence of the expulsion free energy on the core block length and the Flory-Huggins parameter. 
Next, we introduce the end-to-end distance of the core block as a second reaction coordinate and construct a two-dimensional free energy landscape. Using the string method, we identify the minimum energy path and elucidate the chain conformation at the transition state. 
Finally, we conclude in Section \ref{sec:summary} with a brief summary and suggest possible extensions.

%%%%%%%%%%%%%%%%%%%%%%%%%%%%%%%%%%%%%%
\section{Computational Methods}
\label{sec:method}

\subsection{Self-Consistent Field Theory}
We consider a system consisting of $n_\mathrm{P}$ number of AB diblock chains and $n_\mathrm{S}$ number of short homopolymer chains as solvents in a volume $V$.
Each diblock chain comprises $N_\mathrm{A}=f_\mathrm{A}N$ A-segments and $N_\mathrm{B}=f_\mathrm{B}N$ B-segments. 
We independently vary $f_\mathrm{A}$ or $f_\mathrm{B}$ to study the effect of block length,
%Modifying one block length while keeping the other constant changes the total degree of polymerization. 
therefore $f_\mathrm{A} + f_\mathrm{B}$ does not usually equal 1. 
In this work, the total degree of polymerization $(f_{\mathrm{A}}+f_{\mathrm{B}})N$ ranges from $0.8 N$ to $1.5 N$, where $N$ is a reference number of Kuhn segments. 
We focus on star-like micelles with $f_\mathrm{A} > f_\mathrm{B}$, and adopt $\mathrm{A}_{0.8}\mathrm{B}_{0.4}$ as a representative system.
In this parameter range, the corona is relatively dense. 
This places our system within the dense corona regime where the hyperstretching mechanism was observed in recent DPD simulations \cite{Seeger2021, Seeger2022}.
The Flory-Huggins interaction parameters were set to $\chi_{\mathrm{BS}} N = \chi_{\mathrm{AB}} N=40$, $\chi_{\mathrm{AS}} N=0$.
We vary the interaction parameter $\Delta \chi N = \chi_{\mathrm{BS}} N - \chi_{\mathrm{AS}}N$ (with $\chi_{\mathrm{AB}} N = \chi_{\mathrm{BS}} N$) when examining the solvent selectivity.
The solvent chains consist of $N_\mathrm{S}=f_\mathrm{S}N$ segments.
We employ a fixed $f_\mathrm{S}=0.1$ for all systems.
All segments are assumed to have equal Kuhn length $b$ and number density $\rho_0$.
We perform the calculation in the grand canonical ensemble.
The polymer concentration $\phi_\mathrm{P}$ depends on the chemical potentials of polymer and solvent, $\mu_\mathrm{P}$ and $\mu_\mathrm{S}$.
Since the system is incompressible, the two chemical potentials are not independent, and we set $\mu_\mathrm{S} = 0$.
Accordingly, $\phi_\mathrm{P}$ is regulated by $\mu_\mathrm{P}$ or the activity $z_\mathrm{P} = \exp(\mu_\mathrm{P} / k_\mathrm{B} T)$, where $k_\mathrm{B} T$ sets our energy unit.

Within the approximations of the mean-field treatment and Gaussian-chain model, the grand free energy $\mathcal{G}$ of the AB/S system is expressed as 
\begin{align}
  G = \frac{N \mathcal{G}}{k_\mathrm{B} T \rho_0 V}  = &-\frac{Q_\mathrm{S}}{f_{\mathrm{S}}} - \frac{z_\mathrm{P}Q_\mathrm{P}}{f_{\mathrm{A}}+f_{\mathrm{B}}} + \frac{1}{V} \int \mathrm{d} \mathbf{r} \Big\{ \chi_\mathrm{AB} N \phi_\mathrm{A}(\mathbf{r}) \phi_\mathrm{B}(\mathbf{r})   \nonumber \\ 
  & + \chi_\mathrm{AS}N\phi_\mathrm{A}(\mathbf{r})\phi_\mathrm{S}(\mathbf{r}) + \chi_\mathrm{BS}N\phi_\mathrm{B}(\mathbf{r})\phi_\mathrm{S}(\mathbf{r}) \nonumber \\
  & -\omega_\mathrm{A}(\mathbf{r})\phi_\mathrm{A}(\mathbf{r}) - \omega_\mathrm{B}(\mathbf{r})\phi_\mathrm{B}(\mathbf{r}) - \omega_\mathrm{S}(\mathbf{r})\phi_\mathrm{S}(\mathbf{r}) \nonumber \\
  & - \eta(\mathbf{r})[1-\phi_\mathrm{A}(\mathbf{r}) - \phi_\mathrm{B}(\mathbf{r}) - \phi_\mathrm{S}(\mathbf{r})] \Big\}
\end{align}
where $\phi_\mathrm{K}(\mathbf{r})$ (K = A,B,S) is the volume fraction of the K-component and $\omega_\mathrm{K}(\mathbf{r})$ is the conjugate chemical potential field.
$\eta(\mathbf{r})$ is the Lagrange multiplier to enforce the incompressibility.
$Q_\mathrm{P}$ and $Q_\mathrm{S}$ are the single-chain partition functions
\begin{align}
  Q_\mathrm{P} &= \frac{1}{V} \int \mathrm{d} \mathbf{r} \, q_\mathrm{P}(\mathbf{r}, s) q_\mathrm{P}^{\dag}(\mathbf{r}, s) \\
  Q_\mathrm{S} &= \frac{1}{V} \int \mathrm{d} \mathbf{r} \, q_\mathrm{S}(\mathbf{r}, s) q_\mathrm{S}(\mathbf{r}, f_{\mathrm{S}}-s) 
\end{align}
where $q(\mathbf{r}, s)$ and $q^{\dag}(\mathbf{r}, s)$ are the end-integrated propagators. 
For diblocks, they satisfy the modified diffusion equations 
\begin{align}
  \frac{\partial q_\mathrm{P}(\mathbf{r}, s )} {\partial s} &= \frac{Nb^2}{6}\nabla^{2} q_\mathrm{P}(\mathbf{r}, s) - \omega(\mathbf{r}, s) q_\mathrm{P}(\mathbf{r}, s) \\
  - \frac{\partial q_\mathrm{P}^{ \dag } (\mathbf{r}, s )} {\partial s} &= \frac{Nb^2}{6}\nabla^{2} q_\mathrm{P}^{\dag} (\mathbf{r}, s) - \omega(\mathbf{r}, s) q_\mathrm{P}^{\dag} (\mathbf{r}, s)
\end{align}
where $\omega(\mathbf{r}, s) = \omega_\mathrm{K}(\mathbf{r})$ when the $s$-th segment belongs to the K-block.
There is only one type of end-integrated propagator $q_\mathrm{S}(\mathbf{r}, s)$ for the solvents, satisfying the following modified diffusion equation
\begin{align}
  \frac{\partial q_\mathrm{S}(\mathbf{r}, s )} {\partial s} &= \frac{Nb^2}{6}\nabla^{2} q_\mathrm{S}(\mathbf{r}, s) - \omega_\mathrm{S}(\mathbf{r}) q_\mathrm{S}(\mathbf{r}, s) 
\end{align}

In all our calculations, $R_\mathrm{g} = \sqrt{Nb^2 / 6}$ is chosen as the length unit. 
The reference $N$ is chosen as the unit of the contour length, thus $s \in [0, f_\mathrm{A} + f_\mathrm{B}]$ for AB chain and $s \in [0, f_{\mathrm{S}}]$ for solvents. 
The initial conditions of the propagator functions are $q_\mathrm{P}(\mathbf{r}, 0) = q_\mathrm{S}(\mathbf{r}, 0) = 1$ and $q_\mathrm{P}^{\dag}(\mathbf{r}, f_\mathrm{A} + f_\mathrm{B}) = 1$.
Minimization of the free energy functional with respect to $\phi_\mathrm{K}(\mathbf{r})$ and $\omega_{\mathrm{K}}(\mathbf{r})$ leads to the following self-consistent equations
\begin{align}
  \omega_\mathrm{A} (\mathbf{r}) &= \chi_{\mathrm{AB}}N \phi_\mathrm{B}(\mathbf{r}) + \chi_{\mathrm{AS}}N \phi_\mathrm{S}(\mathbf{r}) + \eta(\mathbf{r}) \\
  \omega_\mathrm{B} (\mathbf{r}) &= \chi_{\mathrm{AB}}N \phi_\mathrm{A}(\mathbf{r}) + \chi_{\mathrm{BS}}N \phi_\mathrm{S}(\mathbf{r}) + \eta(\mathbf{r}) \\
  \omega_\mathrm{S} (\mathbf{r}) &= \chi_{\mathrm{AS}}N \phi_\mathrm{A}(\mathbf{r}) + \chi_{\mathrm{BS}}N \phi_\mathrm{B}(\mathbf{r}) + \eta(\mathbf{r}) \\
  \phi_\mathrm{A} (\mathbf{r}) &= \frac{z_\mathrm{P}}{f_{\mathrm{A}}+f_{\mathrm{B}}} \int_{0}^{f_\mathrm{A}} \mathrm{d}s \, q_\mathrm{P}(\mathbf{r}, s) q_\mathrm{P}^{\dag}(\mathbf{r}, s) \\
  \phi_\mathrm{B} (\mathbf{r}) &= \frac{z_\mathrm{P}}{f_{\mathrm{A}}+f_{\mathrm{B}}} \int_{f_\mathrm{A}}^{f_\mathrm{A}+f_\mathrm{B}} \mathrm{d}s\, q_\mathrm{P}(\mathbf{r}, s) q_\mathrm{P}^{\dag}(\mathbf{r}, s) \\
  \phi_\mathrm{S} (\mathbf{r}) &= \frac{1}{f_{\mathrm{S}}} \int_{0}^{f_\mathrm{S}} \mathrm{d}s \, q_\mathrm{S}(\mathbf{r}, s) q_\mathrm{S}(\mathbf{r}, f_{\mathrm{S}}-s) 
\end{align}

These equations can be solved numerically using standard methods.
The modified diffusion equations are solved using pseudospectral method \cite{Rasmussen2002, Tzeremes2002}, with Anderson mixing \cite{Thompson2004, Stasiak2011} was implemented to accelerate convergence. 
Convergence was achieved when the changes in both the field and the incompressibility condition between two successive iterations were less than $10^{-7}$.
We use a lattice of $128\times128\times128$ for spatial discretization and a cubic box of side length $15 R_\mathrm{g}$.
The grid spacing is about $0.117 R_\mathrm{g}$.
The discretization of all polymer chain contours is performed based on a fixed reference step size $\Delta s = 0.01$.
%For a chain with a total degree of polymerization of $(f_\mathrm{A} + f_\mathrm{B})N$, the number of discrete points along its contour is given by $100(f_\mathrm{A} + f_\mathrm{B})$.

%==================================================
\subsection{Single reaction coordinate} 

To quantify the extent of single chain expulsion, we choose a reaction coordinate $r_\mathrm{j} = | \mathbf{r}_{\mathrm{j}} - \mathbf{r}_{\mathrm{cm}}|$ to be the distance between the A-B junction $\mathbf{r}_{\mathrm{j}}$ and the micelle center-of-mass $\mathbf{r}_{\mathrm{cm}}$.
Since the expulsion of a single chain occurs much faster than the structural relaxation of the micelle, we therefore treat the chemical potential field as frozen during this process. 
Under this assumption, the single-chain free energy landscape along the expulsion pathway simplifies to computing the free energy of a single chain at different positions $\mathbf{r}_\mathrm{j}$ under fixed chemical potential fields.

\begin{figure}[htbp]
  \includegraphics[width=1.0\columnwidth]{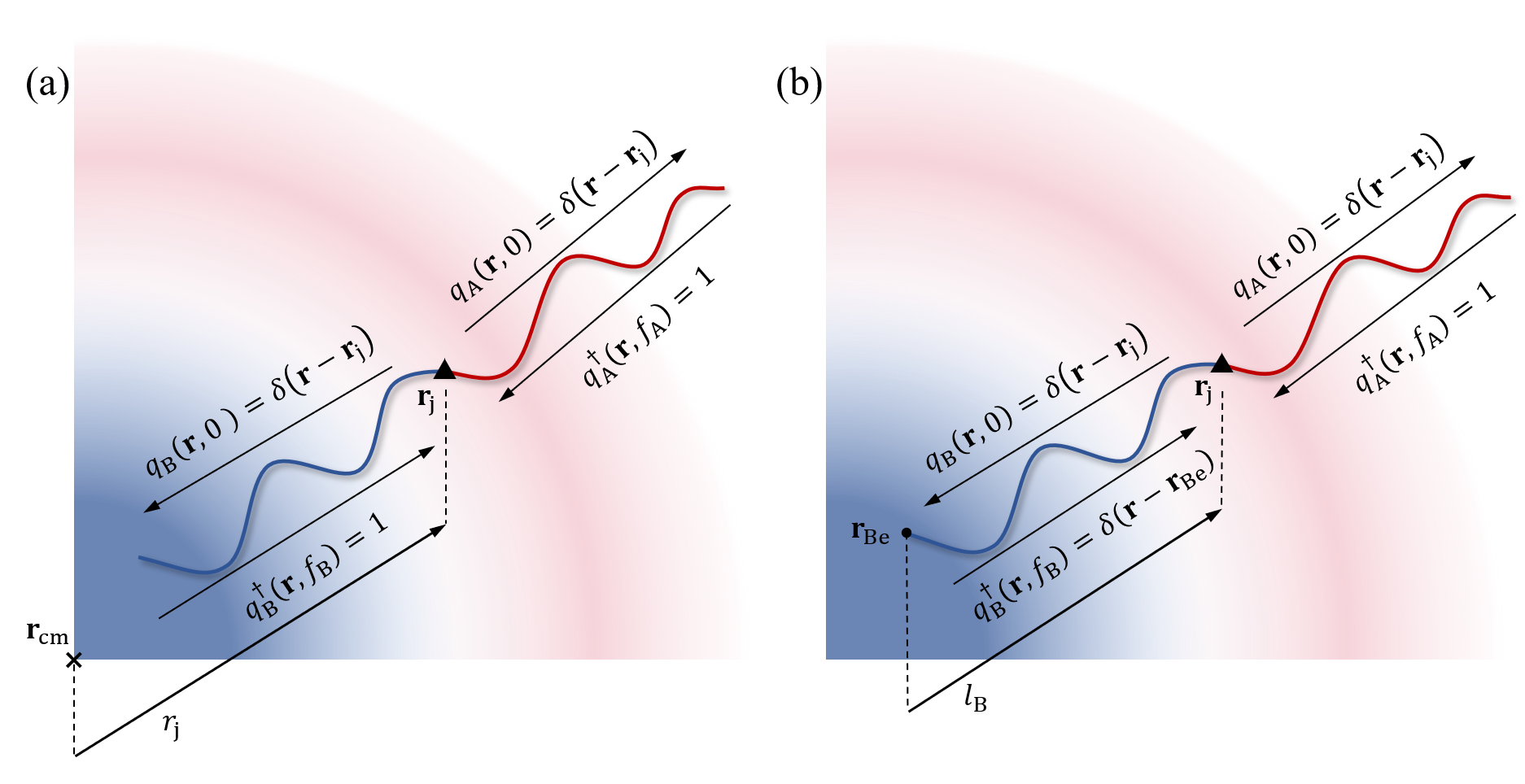}
  \caption{
    Schematic of the propagator calculation for a single chain constrained along specified reaction coordinates in a micelle.
    The blue and red solid lines represent the B (core) and A (corona) blocks, respectively.
    The A-B junction point $\mathbf{r}_{\mathrm{j}}$ is marked by a triangle.  
    The center-of-mass of the micelle $\mathbf{r}_{\mathrm{cm}}$ is marked by a cross.  
    The end of the hydrophobic block $\mathbf{r}_{\mathrm{Be}}$ is marked by a black dot.
    (a) Configuration for one-dimensional free energy profile.
    The reaction  coordinate is the distance between the junction and the center-of-mass of micelle, $r_\mathrm{j}$.
    (b) Extension to two-dimensional free energy landscape. 
    The second coordinate is the distance between the end of the hydrophobic block and the junction point, $l_\mathrm{B}$.
    }
  \label{fig2} 
\end{figure}

Within this framework, the modified diffusion equation is solved with different initial conditions to obtain the partition function $Q_\mathrm{P}[\omega; \mathbf{r}_\mathrm{j}]$ of a single chain with junction point constrained at $\mathbf{r}_\mathrm{j}$ [Figure \ref{fig2}(a)].
In the standard SCFT calculation, the propagator $q_{\mathrm{P}}$ is used for the entire diblock chain. 
For a single-chain calculation where the junction point is constrained, the conformations of the A and B blocks are independent.
The partition function of the entire chain can be factorized as $Q_\mathrm{P}[\omega; \mathbf{r}_\mathrm{j}] = Q_\mathrm{A}[\omega; \mathbf{r}_\mathrm{j}] Q_\mathrm{B}[\omega; \mathbf{r}_\mathrm{j}]$.
The propagator is split into two parts, $q_{\mathrm{A}}$ and $q_{\mathrm{B}}$, for the A and B blocks, respectively.

The initial conditions for the fixed A-B junction are $q_{\mathrm{A}}(\mathbf{r},0) = q_{\mathrm{B}}(\mathbf{r},0) = \delta(\mathbf{r}-\mathbf{r}_\mathrm{j})$, where $\delta(\mathbf{r})$ is Dirac delta function. 
The initial conditions for the free ends remain the same $ q_\mathrm{A}^{\dag}(\mathbf{r},f_\mathrm{A}) = q_\mathrm{B}^{\dag}(\mathbf{r},f_\mathrm{B}) = 1$.

For a given reaction coordinate $\mathbf{r}_\mathrm{j}$, the free energy of the chain is $F(\mathbf{r}_\mathrm{j}) = - \ln{Q_\mathrm{P}[\omega; \mathbf{r}_\mathrm{j}]} $.
We choose the ground state at $|\mathbf{r}_\mathrm{j}|=R_{\mathrm{core}}$ where the free energy takes the minimum, then the free energy difference is 
\begin{align}
  \Delta F(\mathbf{r}_\mathrm{j}) =  F(\mathbf{r}_\mathrm{j})- F(R_\mathrm{core}) = - \ln \left( {\frac{Q_\mathrm{P}[\omega; \mathbf{r}_\mathrm{j}]}{Q_\mathrm{P}[\omega; R_{\mathrm{core}}]}} \right) 
\end{align}
$\Delta F(\mathbf{r}_\mathrm{j})$ characterizes the free energy associated with displacing a single chain to a given location $\mathbf{r}_\mathrm{j}$.

%============================ 2D
\subsection{Two reaction coordinates}
 
In order to capture the hydrophobic block conformation during chain expulsion, we further introduced a second reaction coordinate $l_\mathrm{B}$.
This coordinate is defined as the distance between the end of the hydrophobic B block $\mathbf{r}_{\mathrm{Be}}$ and the A-B junction point $\mathbf{r}_{\mathrm{j}}$.
We consider only the case where the hydrophobic end is aligned along the line connecting the micelle center-of-mass to the copolymer junction point.
When hydrophobic end lies closer to the micelle center than the junction point, $l_\mathrm{B} > 0$; otherwise, it is negative.
The magnitude of $l_{\mathrm{B}}$ indicates the degree of extension of B-block. 
In this case, the boundary condition for B-end is modified as $q_{\mathrm{B}}^{\dagger}(\mathbf{r}, f_{\mathrm{B}}) = \delta(\mathbf{r} - \mathbf{r}_{\mathrm{Be}})$ [Figure \ref{fig2}(b)].

%============================
\subsection{String Method}
We employ the string method \cite{Weinan2002,Weinan2007} to compute the minimum energy path (MEP) connecting the reference state basin and the expulsion basin on the 2D potential energy surface $F(x,y)$.
The initial path is discretized into $N+1$ beads, $\{\boldsymbol{\varphi}_i^0, i=0,1,\dots,N\}$, and connects the basin of the reference state to that of the expulsion state.
In this study, we use $50$ beads for the calculations.
Multiple initial paths with different stretching extents are used to avoid convergence to metastable states.

Subsequently, the following steps are iterated until convergence:
\begin{enumerate}
 \item The first step involves the evolution of the beads, where each bead is moved according to the full potential force field:
  \[
  \boldsymbol{\varphi}_i^* = \boldsymbol{\varphi}_i^n - \Delta t \nabla V(\boldsymbol{\varphi}_i^n)
  \]
  Here, $n$ denotes the iteration step (with $n=0$ representing the initial path), and $\Delta t$ is the time step in the iteration. 
  This ordinary differential equation is solved using the forward Euler method.

  \item The second step is the reparameterization of the string.
  We parameterize by equal arc length to maintain a uniform distribution of beads along the path.
  The cumulative arc length of the current path is first computed: $s_0 = 0$, $s_i = s_{i-1} + |\boldsymbol{\varphi}_i^* - \boldsymbol{\varphi}_{i-1}^*|$, where $i = 1, \dots, N$. 
  The arc length parameters are then normalized: $\alpha_i^* = s_i / s_N$.
  Cubic spline interpolation is used to map the bead positions from the non-uniform grid $\{\alpha_i^*\}$ to the uniform grid $\{\alpha_i = i/N\}$, updating the bead positions $\boldsymbol{\varphi}_i^{n+1}$ for the next iteration.
\end{enumerate}

The above steps are repeated until convergence is achieved. 
Convergence is judged by computing the maximum displacement $d$ of the beads between consecutive iterations:
\[
d = \max_i \frac{|\boldsymbol{\varphi}_i^{n+1} - \boldsymbol{\varphi}_i^n|}{\Delta t}
\]
The iteration will stop when $d$ is less than a preset tolerance $\mathrm{TOL}$.

We set the convergence criterion to $\mathrm{TOL} < 0.0005$, under which the numerical fluctuation of the barrier is less than $0.01 k_\mathrm{B}T$.
Given that chain expulsion is primarily driven by thermal fluctuations, and typical expulsion barriers are on the order of $10\sim20 k_\mathrm{B}T$, excessively high convergence precision is of limited physical relevance.
The potential energy surface in the transition state region is relatively flat, the convergence behavior is sensitive to the step size (Supplementary Material Sec. A).
Despite this sensitivity, the boundaries of the reaction channel and the near-degeneracy of the barriers are robust across the tested range of step sizes.
We chose a step size of $0.02$ to obtain stable results.

%%%%%%%%%%%%%%%%%%%%%%%%%%%%
\section{Results and Discussions}
\label{sec:result}

%==========================================
\subsection{Thermodynamic Stability of Micelles}

\begin{figure}[htbp]
  \includegraphics[width=1.0\columnwidth]{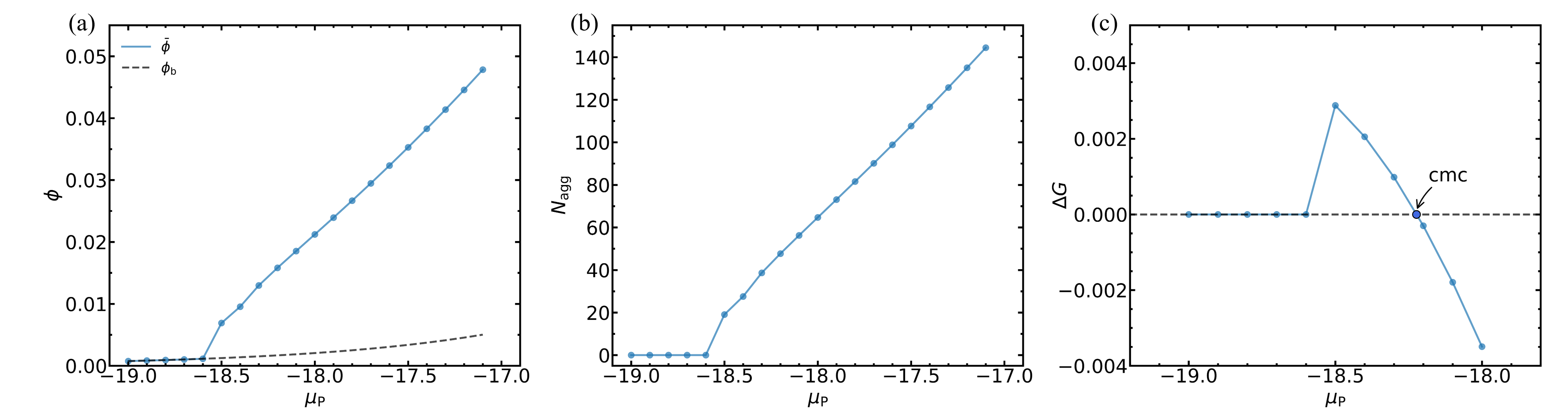}
  \caption{
    (a) Bulk concentration $\phi_{\mathrm{b}}$ and total concentration $\bar{\phi}$ as a function of chemical potential in the polymer solution.
    (b) The aggregation number of micelles $N_\mathrm{agg}$ with different chemical potentials $\mu _\mathrm{P}$.
    (c) Free energy difference $\Delta G$ between micelle solution and disordered solutions, plotted as a function of the chemical potential of polymer $\mu _\mathrm{P}$.
    The parameters of the system are $f_\mathrm{A}=0.80, f_\mathrm{B}=0.40$, $\Delta \chi N=40$.
    } 
  \label{fig3} 
\end{figure}

In the grand canonical ensemble, the size of micelles is tuned by the chemical potential of the chains $\mu_\mathrm{P}$.
%Higher $\mu_\mathrm{P}$ corresponds to a higher polymer concentration. 
We calculate the total concentration $\bar{\phi}$ and the bulk concentration $\phi_{\mathrm{b}}$ for different chemical potentials. 
Figure \ref{fig3}(a) shows the results for the representative system, with $f_\mathrm{A}=0.8, f_\mathrm{B}=0.4$, $\Delta \chi N =40$, and $\chi_{\mathrm{AS}} N =0$. 
When the chemical potential exceeds a critical value (around $\mu_\mathrm{P}=-18.5$), polymer chains begin to aggregate. 
The total concentration $\bar{\phi}$ becomes higher than the bulk concentration $\phi_{\mathrm{b}}$, and the excess portion begins to form micelles.

To quantitatively characterize the size of the micelle, we calculate the aggregation number  $N_\mathrm{agg}$ from the polymer concentration profile $\phi(\mathbf{r})$. 
This approach is adapted from the method of Ref. \cite{Zhou2011}, with the modification that we include the bulk concentration $\phi_{\mathrm{b}}$ to account for the finite background polymer density in our system.
The micelle is considered as a localized region of polymer enrichment. 
%To determine its size without arbitrarily defining a boundary, we calculate it based on mass conservation. 
The total amount of polymer in the system, $\int_V \phi(\mathbf{r}) \mathrm{d}\mathbf{r}$, is the sum of the amount in the micelle and that in the bulk solution: 
$\phi_{\mathrm{micelle}} V_{\mathrm{micelle}} + \phi_{\mathrm{b}} (V-V_{\mathrm{micelle}}) = \int _V \phi (\mathbf{r}) d \mathbf{r}$.
Here, $\phi_{\mathrm{b}}$ is the bulk concentration, obtained by averaging the concentration profile near the box boundary (the region far from the micelle).
$\phi_{\mathrm{micelle}}$ is the average polymer concentration inside the micelle. 
Since our system is in the strong segregation regime, the micelle core is nearly solvent-free, so $\phi_{\mathrm{micelle}} \approx 1$. %(pure polymer volume fraction).
From the above equation, we obtain the expression for the micelle volume $V_{\mathrm{micelle}}$:
\begin{align}
V_{\mathrm{micelle}} = \frac{\int_V [\phi (\mathbf{r}) - \phi_{\mathrm{b}}] d \mathbf{r}} {1 - \phi_{\mathrm{b}}}.
\end{align}
The aggregation number is then given by
\begin{align}
  N_\mathrm{agg}=\frac{\rho_0 \, V_\mathrm{micelle} } {(f_\mathrm{A}+f_\mathrm{B})N} \, .
\end{align}
The aggregation number of the micelle increases with chemical potential $\mu_\mathrm{P}$, as shown in Figure \ref{fig3}(b).
In subsequent studies, we keep the aggregation number constant while varying other parameters such as the block lengths $f_\mathrm{A}$, $f_\mathrm{B}$, and the solvent selectivity $\Delta \chi N$.

The energy difference between the homogeneous solution and the micelle, $\Delta G$,  is shown in Figure \ref{fig3}(c).
After micelle formation, as the chemical potential increases, the micelles grow and the aggregation number increases. 
Initially, the free energy of the micelle system exceeds that of the disordered solution, $\Delta G > 0$, because the micelles are metastable. 
At a specific chemical potential, the free energy is equal to that of the disordered phase, $\Delta G = 0$, and the corresponding copolymer concentration is defined as the critical micelle concentration (CMC). 
Beyond this chemical potential, the free energy of micelles becomes lower than that of the disordered phase, $\Delta G < 0$, and the micelles become thermodynamically stable. 
All systems studied in this work are with the diblock concentration above CMC.

%===================================
\subsection{One Reaction Coordinate}

\begin{figure}[htbp]
  \includegraphics[width=1.0\columnwidth]{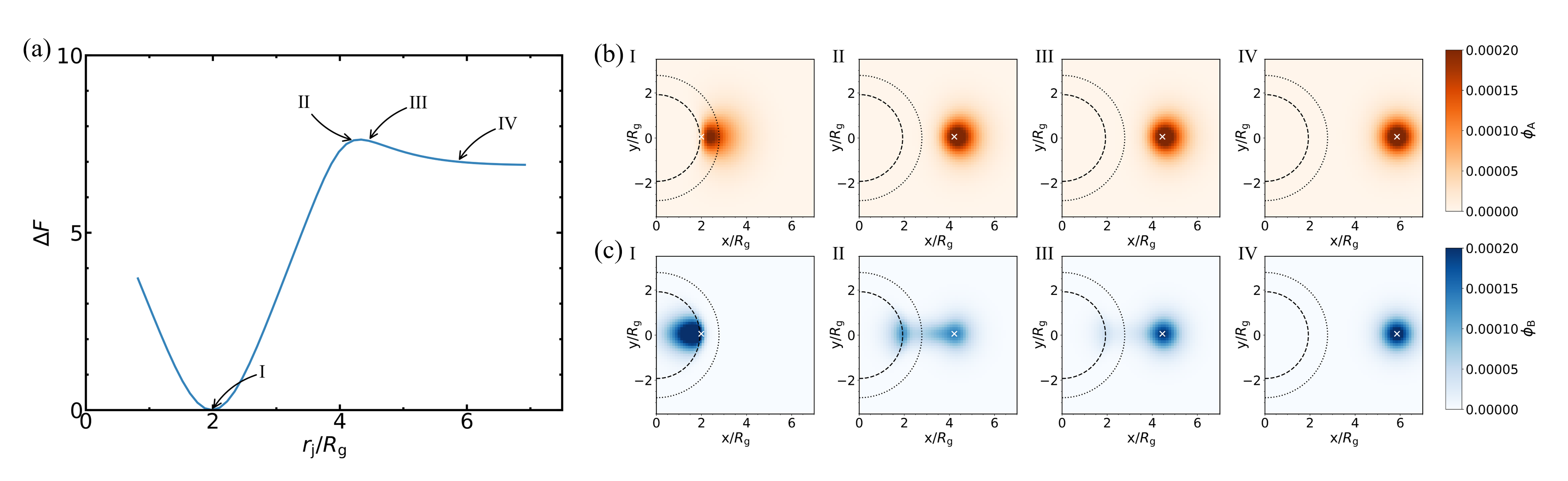}
  \caption{
    (a) The free energy profiles of chain expulsion from a micelle (with $f_\mathrm{A}=0.8, f_\mathrm{B}=0.4, \Delta \chi N=40, N_\mathrm{agg}=100$) as a function of the distance between the center-of-mass of the micelle and the A-B junction ($r_\mathrm{j}$).
    The orange heatmap (b) and blue heatmap (c) show the density distribution of the A and B blocks on the cross-sectional plane at $Lz/2$.
    Points I to IV represent different states during the expulsion process.  
    Point I ($r_{\mathrm{j}} = 2.0 R_{\mathrm{g}}$) represents the reference state.  
    Point II ($r_{\mathrm{j}} = 4.2 R_{\mathrm{g}}$) and Point III ($r_{\mathrm{j}} = 4.5 R_{\mathrm{g}}$) represent the state before and after the transition state, respectively.
    Point IV ($r_{\mathrm{j}} = 5.9 R_{\mathrm{g}}$) represents the fully expelled state.
    }
  \label{fig4}
\end{figure}

Figure \ref{fig4}(a) presents the free energy profile for the single chain expulsion from a micelle of $N_\mathrm{agg} =  100 $ in the representative system.
The free energy profile exhibits a minimum, corresponding to the equilibrium state where the junction resides at the core-corona interface. 
This state is chosen as the reference for the expulsion process, and the reaction coordinate corresponds to the core radius $R_\mathrm{core}$ of the micelle. 
The micelle core radius $R_\mathrm{core}$ and the overall micelle radius $R_\mathrm{micelle}$ were determined from the micelle volume, yielding $R_\mathrm{micelle} = (3V_\mathrm{micelle}/4\pi)^{1/3} = 2.9 R_\mathrm{g}$ and $R_\mathrm{core} = [3V_\mathrm{micelle} f_\mathrm{B}/4\pi(f_\mathrm{A}+f_\mathrm{B})]^{1/3} = 2.0 R_\mathrm{g}$. 

As the reaction coordinate increases, the hydrophobic block is progressively drawn out of the core, becoming exposed to the solvent and corona chains. 
The free energy increases because of unfavorable enthalpic contributions. 
The maximum in the free energy profile identifies the transition state, which in this system coincides with the extraction of most of the hydrophobic segment from the micelle core. 
The free energy difference $ \Delta F $ between this transition state and the reference state defines the expulsion barrier $F_\mathrm{expel}$. 
Once the chain crosses the energy barrier, it escapes into the solution. 
%The free energy decreases slightly because the chain gains greater conformational freedom as it moves from the crowded corona into the solution.
%This energy difference is measured between the transition state and the isolated chain in solution.
%It corresponds to the insertion barrier $F_\mathrm{insert}$ for the reverse process, \emph{i.e.}, a single chain inserting into the micelle.

The density distributions of the expelled chain are shown in Figure \ref{fig4}(b) and (c) for A and B-blocks. 
When the expelled chain is in the ground state (point I, $r_\mathrm{j} = 2.0 R_\mathrm{g}$), the A and B blocks are localized on opposite sides of the micelle interface. 
During the expulsion process (point II), the B-block retains a stretched density distribution, with a small portion remaining inside the micelle core.
After the transition state (point III, $r_\mathrm{j} = 4.5 R_\mathrm{g}$), the last part of the hydrophobic block moves out of the core.
Once the chain is fully expelled (point IV), the B-block relaxes from its stretched conformation into a more compact state.
These results indicate that the hydrophobic block adopts a stretched conformation during the expulsion process.

\begin{figure}[htbp]
  \includegraphics[width=1.0\columnwidth]{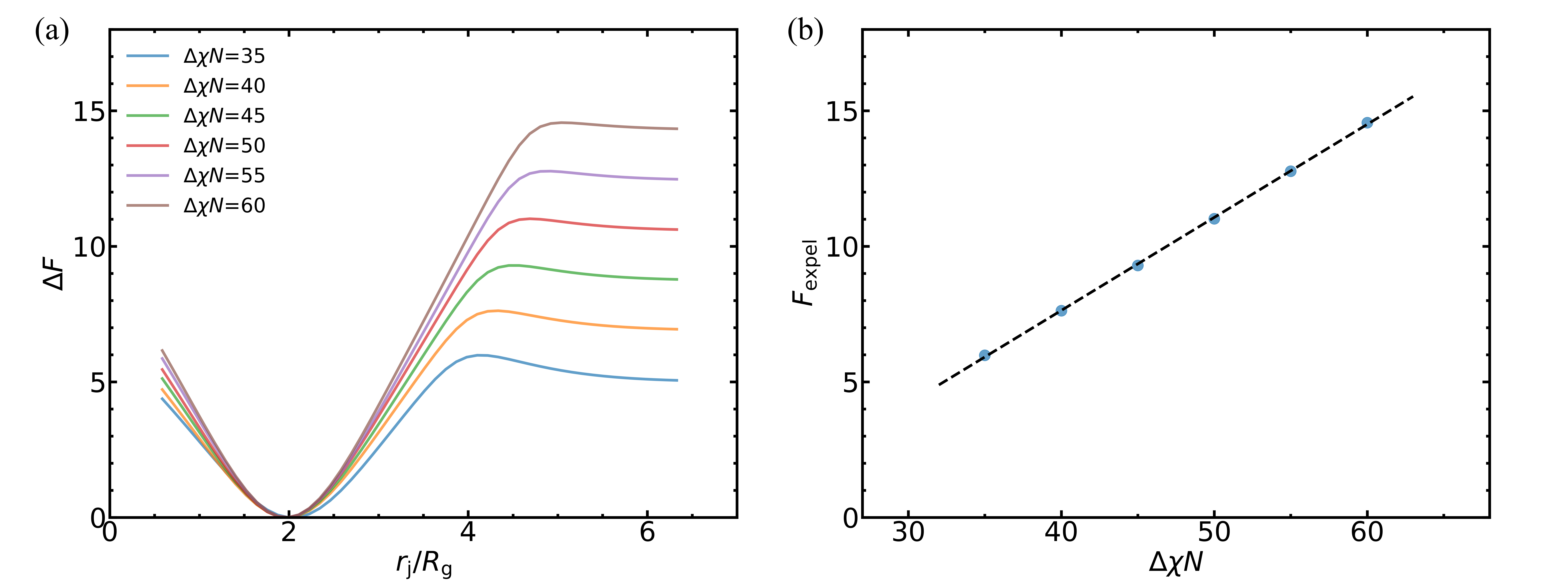}
  \caption{
    Effect of solvent selectivity on chain expulsion free energy.
    The system parameters are $f_\mathrm{A}=0.8, f_\mathrm{B}=0.4, N_\mathrm{agg}=100$.
    (a) Free energy profiles of different $\Delta \chi N$ are plotted as a function of the distance $r/R_\mathrm{g}$.
    (b) Free energy barrier of chain expulsion is plotted as a function of $\Delta \chi N$.
    }
  \label{fig5}
\end{figure}

To investigate the influence of solvent selectivity, quantified by $\Delta \chi N = \chi_{\mathrm{BS}} N - \chi_{\mathrm{AS}}N$, we use the $\mathrm{A}_{0.8}\mathrm{B}_{0.4}$ system as an example. 
The micelle aggregation number was kept at $N_\mathrm{agg}=100$.
The free energy profiles for different $\Delta \chi N$ are shown in Figure \ref{fig5}(a).
The free energy increases more rapidly with larger values of $\Delta \chi N$. 
Near the transition state, $\Delta F$ becomes significantly greater for systems with higher $\Delta \chi N$. 
The contribution of core-solvent incompatibility to the free energy originates from the unfavorable interactions between the core and corona/solvent. 
For a fixed hydrophobic B block length, a larger $\Delta \chi N$ corresponds to stronger hydrophobicity, which leads to a steeper increase in free energy and a higher free energy barrier.
Further analysis reveals a linear relationship between the free energy barrier and $\Delta \chi N$ [Figure \ref{fig5}(b)].
These results are consistent with prior experimental findings \cite{WangEn2020} and molecular dynamics simulations \cite{Seeger2021}.

\begin{figure}[htbp]
  \includegraphics[width=1.0\columnwidth]{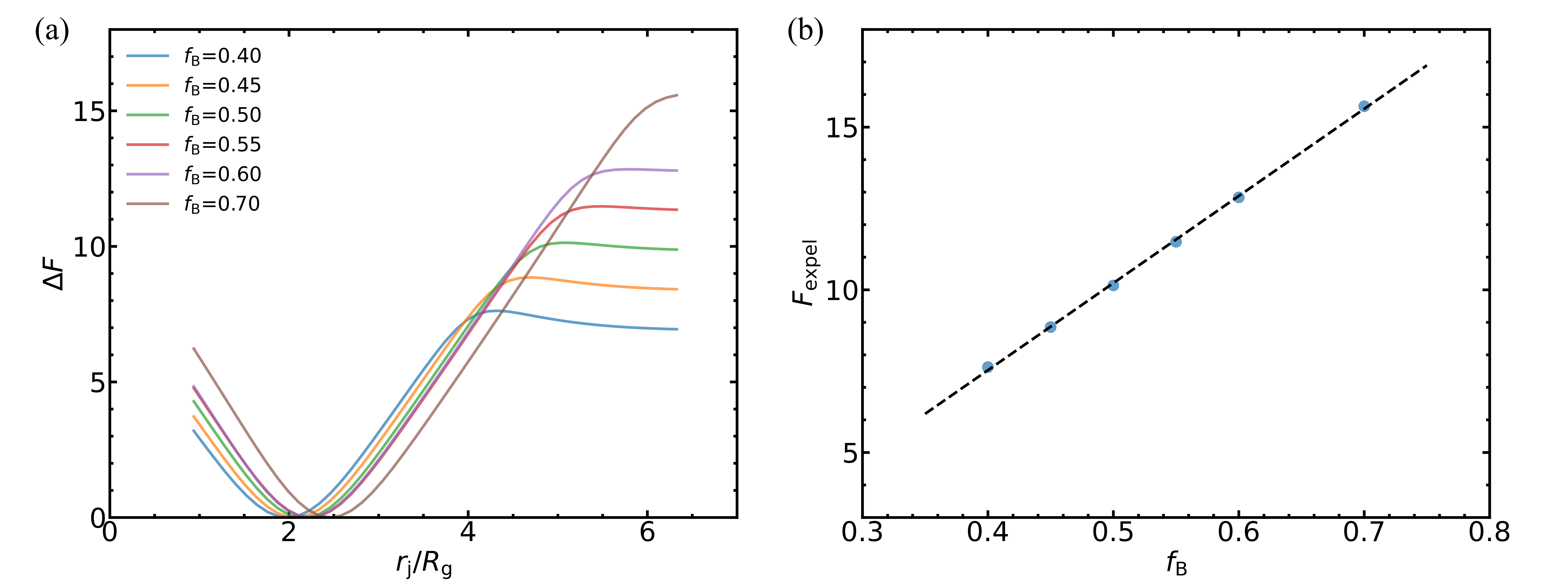}
  \caption{
    Effect of hydrophobic segment length on chain expulsion free energy.
    The system parameters are $f_\mathrm{A}=0.8, \Delta \chi N=40, N_\mathrm{agg}=100$.
    (a) Free energy profiles of different $f_\mathrm{B}$ are plotted as a function of the distance $r/R_\mathrm{g}$.
    (b) Free energy barrier of chain expulsion is plotted as a function of $f_\mathrm{B}$.
    }
  \label{fig6}
\end{figure}

We further investigated the influence of the hydrophobic block length, $f_\mathrm{B}$. 
We employed chains of $\mathrm{A}_{0.8}\mathrm{B}_{x}$ (with $x \in [0.4, 0.7]$), and $N_\mathrm{agg}=100$. 
The interaction parameters were fixed at $\Delta \chi N = 40$.
Free energy profiles for chain expulsion across systems with different hydrophobic block lengths are shown in Figure \ref{fig6}(a). 

The results indicate that when $\Delta \chi N$ is constant, the incremental rate of free energy during the initial stage of chain expulsion remains comparable across different chain lengths.
This is because the enthalpic penalty per unit length for the contact between the hydrophobic block and the solvent is identical. 
Meanwhile, the free energy profile for a longer hydrophobic block exhibits an extended rising period. 
This leads to a higher transition state energy and a greater distance of the transition state from the micelle core. 
This suggests that a longer hydrophobic block results in a larger unfavorable contact area with the solvent or corona during the expulsion process, thereby causing a more significant increase in the free energy of the system.

Further analysis shows that the chain expulsion barrier $F_{\mathrm{expel}}$ scales linearly with the hydrophobic block length, as shown in Figure \ref{fig6}(b). 
The combined scaling relationship $F_\mathrm{expel} \sim f_\mathrm{B} \Delta \chi$ and the density distributions further confirm that the escaping chain adopts a stretched conformation during the process. 
This finding is consistent with the ``bead-by-bead'' expulsion mechanism proposed by Seeger \emph{et al.} \cite{Seeger2022} based on DPD simulations. 
This mechanism suggests that near the transition state, the monomer at the end of the hydrophobic block tends to remain in the micelle core to reduce unfavorable contacts with the solvent, leading to a highly stretched hydrophobic chain. 
Only after crossing the transition state, all hydrophobic segments leave the micelle core.% and the chain subsequently collapses in those simulations.
%Our SCFT results show that near the transition state, the density distribution of the hydrophobic block exhibits a stretched conformation. 
%A low-density region of the hydrophobic block also persists near the micelle core. 
%After expulsion, the hydrophobic block relaxes from this stretched conformation into a more compact state. 
%Although this relaxed state does not represent a true hydrophobic collapse, the stretched conformation we observe at the transition state is consistent with results from particle-based simulations.

Further analysis of the effects of hydrophilic block length and aggregation number is presented in Supplementary Material Sec. B and C, respectively. 
These results show that both effects are small.
We therefore use a fixed aggregation number $N_{\rm agg}=100$ for all systems, which allows us to isolate the effects of chain parameters on the expulsion barrier.

The above results, based on single reaction coordinate SCFT calculations, reproduce the linear scaling relationship of the energy barrier from the perspective of flexible chain theory.
This demonstrates that the first reaction coordinate, $r_\mathrm{j}$, can effectively capture the scaling behavior of the energy barrier during chain expulsion.
However, $r_\mathrm{j}$ is insufficient to distinguish the different conformations of the hydrophobic block.
%More importantly, it cannot directly capture the coupling between the degree of stretching of the hydrophobic block and the chain expulsion process.
Therefore, we introduce the second reaction coordinate  in the next section.
%It will simultaneously constrain the position of the junction point and the terminal end of the hydrophobic block. 
This approach aims to provide a more detailed characterization of the chain's conformational evolution during expulsion and its impact on the free energy landscape.

%==================================
\subsection{Two Reaction Coordinates}

\begin{figure}[htbp]
  \includegraphics[width=0.8\columnwidth]{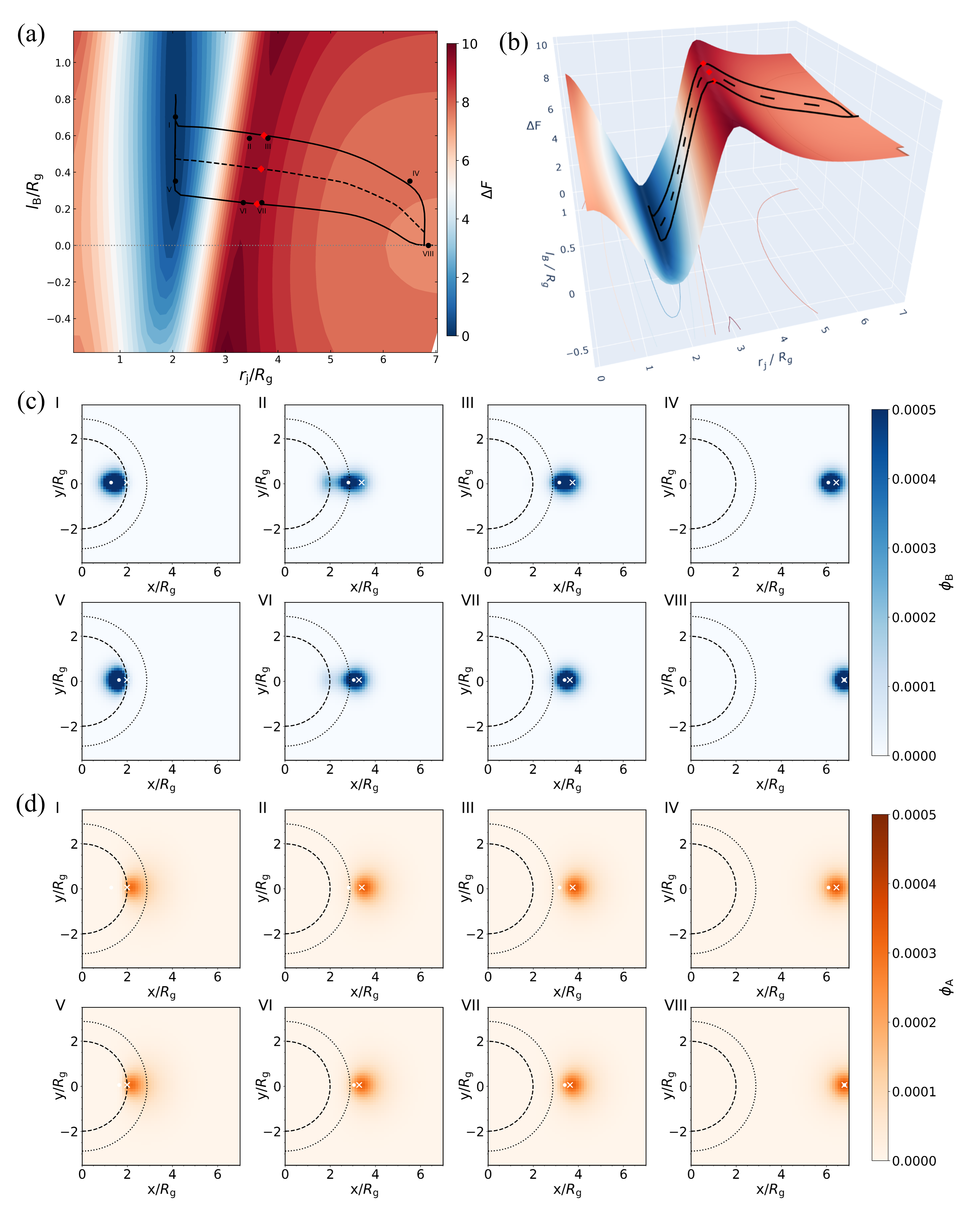}
  \caption{
    The 2D (a) and 3D (b) free energy profiles of chain expulsion from a micelle (with $f_\mathrm{A}=0.8, f_\mathrm{B}=0.4, \Delta \chi N =40, N_\mathrm{agg}=100$).
    The first reaction coordinate is $r_{\mathrm{j}}$ and the second reaction coordinate is $l_{\mathrm{B}}$.
    The black solid line with smaller $l_{\mathrm{B}}$ is MEP1, and the one with larger $l_{\mathrm{B}}$ is MEP2.
    The black dashed line is an intermediate path used for stability testing.
    Red diamonds mark the transition states along the paths.
    The heatmaps show the density distribution of the B-block(c) and A-block(d) on the cross-sectional plane at $Lz/2$.
    Points I to VIII represent different states of the expelled chain during the expulsion process, with $(r_{\mathrm{j}}, l_{\mathrm{B}})$ indicates the reaction coordinates: I (2.0, 0.7), II (3.4, 0.6), III (3.8, 0.6), IV (6.4, 0.4), V (2.0, 0.4), VI (3.3, 0.2), VII (3.6, 0.2), VIII (6.8, 0).
    %    Since the MEP is obtained by interpolation using the string method, the points shown correspond to the closest accessible numerical values along the path.
%    Point I $(2.0, 0.7)$ represents the reference state.
%    Point II $(3.4, 0.6)$ and Point III $(3.8, 0.6)$ are the states before and after the transition state on MEP2, respectively.
%    Point IV $(6.4, 0.4)$ is the state after the transition state on MEP2, where the chain adopts a more compact conformation in solution.
%    Point V $(2.0, 0.4)$ represents a state where the hydrophobic block contracts inside the micelle core during expulsion.
%    Point VI $(3.3, 0.2)$ is the state before the transition state on MEP1.
%    Point VII $(3.6, 0.2)$ is the state just after the transition state on MEP1.
%    Point VIII $(6.8, 0)$ represents the relaxed state in the solution far from the micelle.
    }
  \label{fig7}
\end{figure}

To investigate the energy differences associated with different chain conformations during expulsion, we computed the free energy landscape using two reaction coordinates. 
The first reaction coordinate is the distance between the junction point of the diblock copolymer and the center-of-mass of the micelle (denoted as $r_{\mathrm{j}}$).
This reaction coordinate describes the process of chain expulsion.
The second reaction coordinate is the distance between the hydrophobic end of the polymer chain and the junction point (denoted as $l_{\mathrm{B}}$).
This reaction coordinate is used to track the degree of stretching of the hydrophobic block during the expulsion process. 
Positive $l_{\mathrm{B}}$ indicates the end is closer to the micelle core relative to the junction point; otherwise, it is negative.
The results are shown in Figure \ref{fig7}.

The free energy landscape [Figure \ref{fig7}(b)] illustrates the presence of two valleys of local free energy minima.
The lower valley corresponds to the state where the hydrophobic block remains within the micelle core, and is defined as the reference state valley.
The higher valley corresponds to the state where the chain has been fully expelled, and is defined as the expelled state valley.
The two valleys are separated by a gentle U-shaped ridge, which represents the free energy maximum corresponding to the transition state region.
The energy difference between this ridge and the minimum of the reference state valley gives the free energy barrier that must be overcome during chain expulsion.

The initial path was constructed starting from the minimum of the reference valley [point I, with coordinates $(r_{\mathrm{j}}^{\mathrm{ref}}, l_{\mathrm{B}}^{\mathrm{ref}})$].
It passed through hypothetical transition state positions, a series of points along the ridge with coordinates $(r_{\mathrm{j}}^*, l_{\mathrm{B}}^*)$.
The path ended in the solution far from the micelle [point VIII, with coordinates $(r_{\mathrm{j}}^{\infty}, l_{\mathrm{B}}^{\infty})$].
Here, $r_{\mathrm{j}}^{\mathrm{ref}}$ corresponds to the micelle core radius $R_\mathrm{core}$, where the junction point is located at the interface and the hydrophobic end resides inside the core.
This represents the lowest-energy state of the polymer chain in the micelle.
$l_{\mathrm{B}}^{\mathrm{ref}}$ is the corresponding end-to-end distance of the hydrophobic block in the melt-like dry micelle core.
We define this lowest energy state as the reference state for the expulsion process.

In the solution far from the micelle ($r_{\mathrm{j}}^{\infty}$), the hydrophobic block is exposed to a uniform external field.
The propagator of the hydrophobic block has the same Gaussian spatial form as that of an ideal Gaussian chain.
The end-to-end distribution of the hydrophobic block is therefore governed by entropy.
When the junction point position $\mathbf{r}_{\mathrm{j}}^{\infty}$ is fixed, the probability density of the hydrophobic end $\mathbf{r}_{\mathrm{Be}}$ follows a Gaussian distribution.
The most probable position of this distribution is the junction point itself, which corresponds to the free energy minimum.
The corresponding reaction coordinate is $l_{\mathrm{B}}^{\infty}=|\mathbf{r}_{\mathrm{j}}^{\infty} - \mathbf{r}_{\mathrm{Be}}|= 0$.

Because the U-shaped ridge between the two valleys is very shallow, we constructed multiple initial paths to ensure full coverage of different stretching states of the hydrophobic block.  
These paths were built at intervals of $0.1$ within the range $l_{\mathrm{B}_0}^* / R_\mathrm{g} \in [0, 2]$, and each was used in the string method calculation individually.
The results show that the initial paths do not converge to a single MEP, but instead form a continuous reaction channel.  
We define the lower boundary of this channel, where $l_{\mathrm{B}}$ is smaller, as MEP1 [solid lower line in Figure \ref{fig7}(a)], and the upper boundary as MEP2 [solid upper line in Figure \ref{fig7}(a)].  
The region between them is defined as the reaction channel of the system.

In the MEP2 path, the chain is expelled directly from the reference state (point I) as $r_{\mathrm{j}}$ increases. 
The density distribution heatmaps of the B-block on the cross-sectional plane at $L_z/2$ are shown in Figure \ref{fig7}(c). 
In contrast, the MEP1 path exhibits a state where $l_{\mathrm{B}}$ first decreases (point V), with a density distribution that is closer to the micelle interface compared to point I.  
This process occurs within the reference valley, with little change in free energy.  

Before the transition state (points II and VI), the hydrophobic block exhibits a low density at the core-corona interface.
This conformation combines features of both stretching and compaction, and is neither a budding configuration at the interface nor a completely stretched state. 
Instead, it adopts a lollipop-like conformation, in which most of the expelled portion is concentrated near the junction point while the remaining part is stretched.  
The hydrophobic block conformation tends to localize in the outer layer of the micelle rather than at the core-corona interface, which may be related to the high corona density near the interface that prevents chain compaction.
After the transition state, the chain is expelled into the solution, and the hydrophobic block relaxes from its stretched conformation into a more compact state (points III and VII), until $l_{\mathrm{B}}$ decreases to zero (point IV and VIII).  
Throughout the expulsion process, the degree of stretching of the hydrophobic block has no significant effect on the density distribution of the hydrophilic block [Figure \ref{fig7}(d)].

The energy barrier differences among different paths within the reaction channel are within $\sim 0.05 k_{\mathrm{B}}T$, nearly degenerate.  
This indicates that the end-to-end distance $l_{\mathrm{B}}$ of the hydrophobic block in the transition state region has a distribution range.
The contributions of the paths within the channel to the chain expulsion process are almost identical in terms of the energy barrier.
To further verify the continuity of the reaction channel, we performed an additional string method convergence test using an initial path located midway between MEP1 and MEP2 [dashed line in Figure \ref{fig7}(a)]. 
The results show that this path converges rapidly. 
%This demonstrates that MEP1 and MEP2 are not independent pathways. center-of-mass are just the lower and upper boundaries of a single, continuous, and nearly flat reaction channel. 
The flatness of the free energy surface in this region suggests that the barriers along different paths are comparable. 
The transition state should be viewed as an ensemble of thermally accessible stretched conformations, rather than a single well-defined configuration.

\begin{figure}[htbp]
  \includegraphics[width=0.9\textwidth]{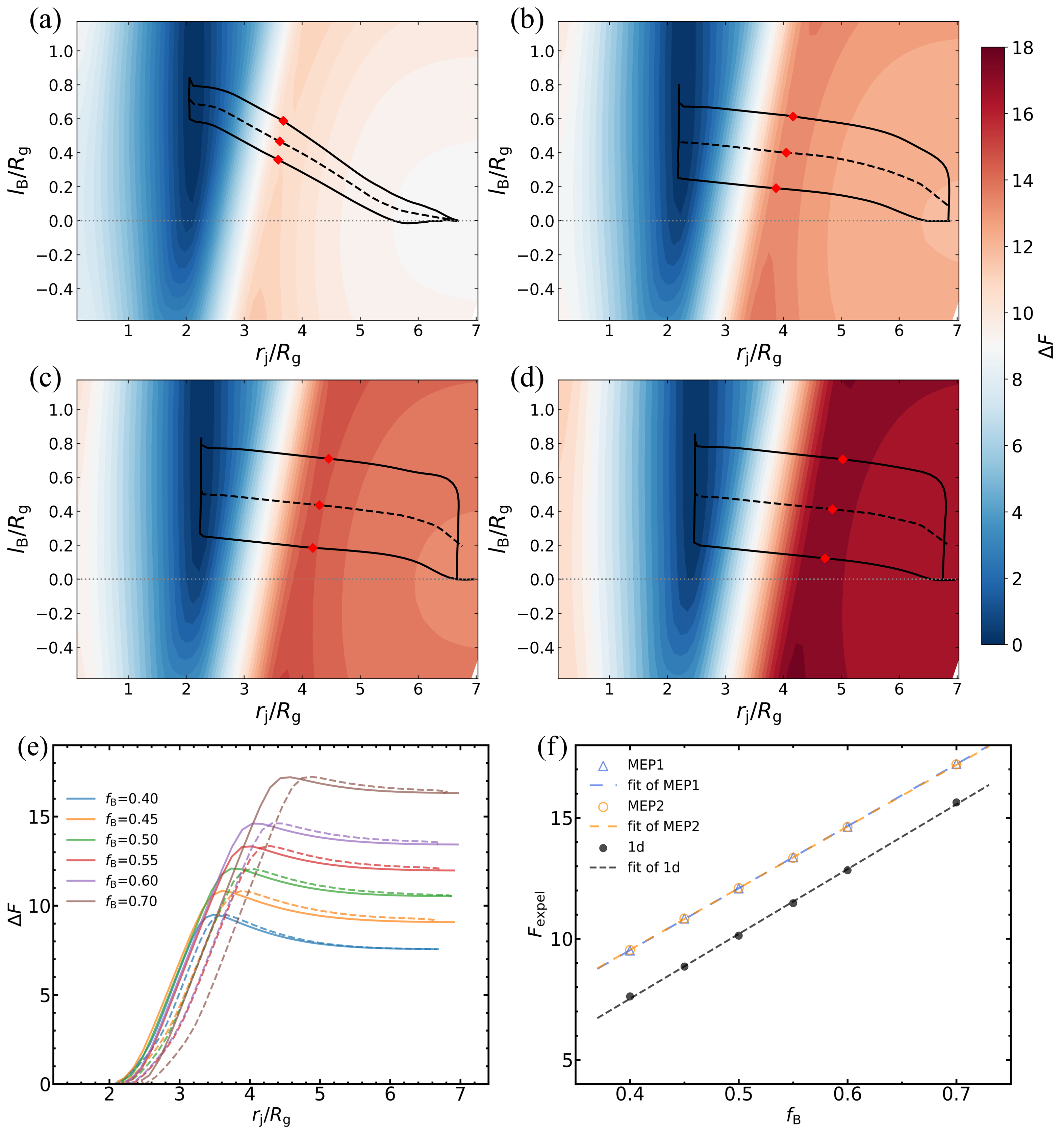}
  \caption{
    Effect of hydrophobic block length on the free energy landscape and minimum energy paths.
    The system parameters are $f_\mathrm{A}=0.8, \Delta \chi N=40, N_{\mathrm{agg}}=100$.
    (a) $f_\mathrm{B}=0.4$, (b) $f_\mathrm{B}=0.5$, (c) $f_\mathrm{B}=0.6$, (d) $f_\mathrm{B}=0.7$.
    The first reaction coordinate is $r_{\mathrm{j}}$ and the second reaction coordinate is $l_{\mathrm{B}}$.
    The black solid line with smaller $l_{\mathrm{B}}$ is MEP1, and the one with larger $l_{\mathrm{B}}$ is MEP2.
    (e) shows the free energy profiles along MEP1 (solid line) and MEP2 (dashed line).
    (f) shows the transition state barriers for MEP1 (blue triangles) and MEP2 (orange circles), along with the corresponding linear fits (blue dashed line and orange dashed line), respectively.
    The black dots and line represent the energy barrier and linear fitting of a 1D system in which only the junction point is constrained.
  }
  \label{fig8}
\end{figure}

We investigated the effects of the hydrophobic block length $f_\mathrm{B}$ and the solvent selectivity $\Delta \chi N$ on the free energy landscape for chain expulsion.
The hydrophobic block length mainly affects the height and position of the ridge and the expelled-state valley, as shown in Figure~\ref{fig8}.
A longer hydrophobic block shifts the expelled-state valley to larger $r_{\mathrm{j}}$, resulting in a larger transition state reaction coordinate $r_{\mathrm{j}}^*$.
Because the solvent selectivity is fixed, the free energy gradient between the reference state valley and the ridge remains nearly unchanged, and the MEP exhibits an almost constant slope, as shown in Figure \ref{fig8}(e).
A longer hydrophobic block reaches a higher-energy transition state by increasing the area of the hydrophobic block exposed to the solvent.

\begin{figure}[htbp]
  \centering
  \begin{subfigure}[b]{0.9\textwidth}
    \includegraphics[width=\textwidth]{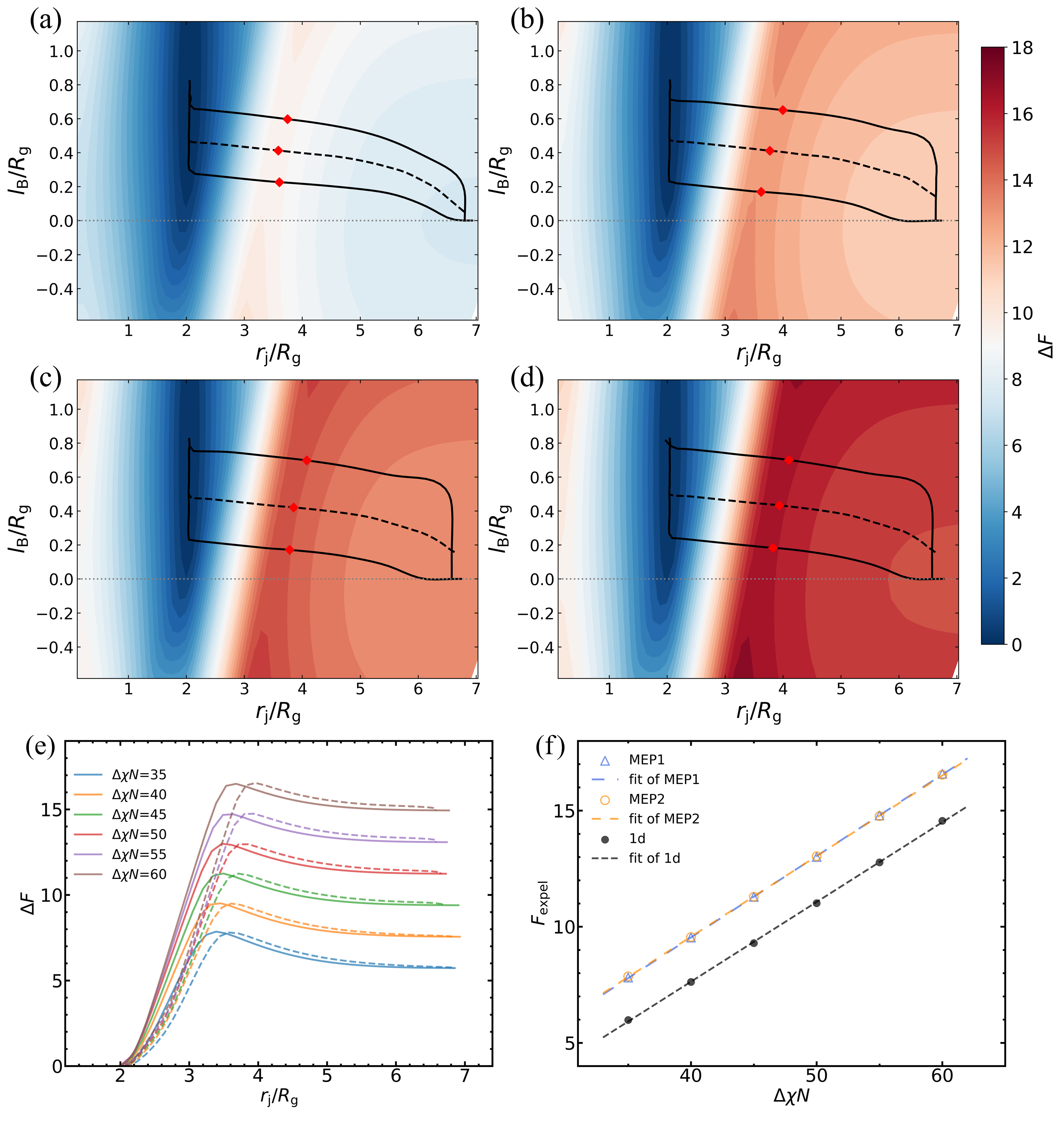}
  \end{subfigure} 
  \caption{
    Effect of solvent selectivity on the free energy landscape and minimum energy paths.
    The system parameters are $f_\mathrm{A}=0.8, f_\mathrm{B}=0.4, N_{\mathrm{agg}}=100$.
    The first reaction coordinate is $r_{\mathrm{j}}$ and the second reaction coordinate is $l_{\mathrm{B}}$.
    The black solid line with smaller $l_{\mathrm{B}}$ is MEP1, and the one with larger $l_{\mathrm{B}}$ is MEP2.
    (a) $\Delta \chi N=40$, (b) $\Delta \chi N =50$, (c) $\Delta \chi N=55$, (d) $\Delta \chi N=60$.
    (e) shows the free energy profiles along MEP1 (solid line) and MEP2 (dashed line).
    (f) shows the transition state barriers for MEP1 (blue triangles) and MEP2 (orange circles), along with the corresponding linear fits (blue dashed line and orange dashed line), respectively.
    The black dots and line represent the energy barrier and linear fitting of a 1D system in which only the junction point is constrained.
  }
  \label{fig9}
\end{figure}

In contrast, the degree of hydrophobicity mainly affects the height of the ridge and the expelled-state valley, with little influence on their positions, as shown in Figure \ref{fig9}.  
As $\Delta \chi N$ increases, the free energy gradient between the reference-state valley and the ridge becomes significantly steeper, requiring the chain to overcome a more sharply rising free energy during expulsion.  
This is reflected in an increasing slope of the MEP with $\Delta \chi N$ [Figure \ref{fig9}(e)], leading to a higher expulsion barrier.  
These results indicate that $f_\mathrm{B}$ and $\Delta \chi N$ influence the transition state barrier through the breadth and intensity of contact between the hydrophobic block and the solvent, respectively.
This is consistent with the behavior observed in our one-dimensional system.

Although $f_\mathrm{B}$ (0.4$\sim$0.7) and $\Delta \chi N$ (35$\sim$60) vary significantly in our systems, the barrier differences between MEP1 and MEP2 in both systems remain within $0.05\,k_\mathrm{B}T$, nearly degenerate [Figure \ref{fig8}(f) and \ref{fig9}(f)].  
The expulsion barriers along the two boundary paths follow the linear scaling relations $F_{\mathrm{expel}} \sim f_\mathrm{B}$ and $F_{\mathrm{expel}} \sim \Delta \chi N$, respectively.  
The differences between them are negligible, indicating that the paths within the channel also obey the corresponding linear relations.

The energy barriers in the 2D system are slightly higher than those in the 1D system. 
In the 1D case, only the junction point is constrained, and the hydrophobic end relaxes freely. 
In the 2D case, the position of the hydrophobic end is additionally constrained, which reduces the conformational entropy and raises the barrier. 
After constraining the end position, the slope of the linear fit to the barrier is nearly unaffected. 
The intercept increases $\sim 2 k_\mathrm{B}T$ and remains nearly constant across both systems. 
This indicates that the loss of conformational entropy caused by the additional end-constraint is largely independent of the hydrophobic block length and the solvent selectivity.

The main factor affecting the width of the reaction channel is not the barrier height, but rather the geometric characteristics of the path from the initial state to the transition state.  
The MEP profiles [Figure \ref{fig8}(e) and \ref{fig9}(e)] show that the most significant difference between MEP1 and MEP2 in both systems occurs in the initial stage.  
MEP1 exhibits a continuously steep rise from the beginning, whereas MEP2 initially shows a more gradual increase in free energy.  
This leads to a difference in the distance traveled from the initial state to the transition state along the two paths.  
The conformational contraction with little free energy change during expulsion in MEP1 may be the microscopic origin of the broadening of the reaction channel.

\begin{figure}[htbp]
  \centering
  \begin{subfigure}[b]{0.99\textwidth}
    \includegraphics[width=\textwidth]{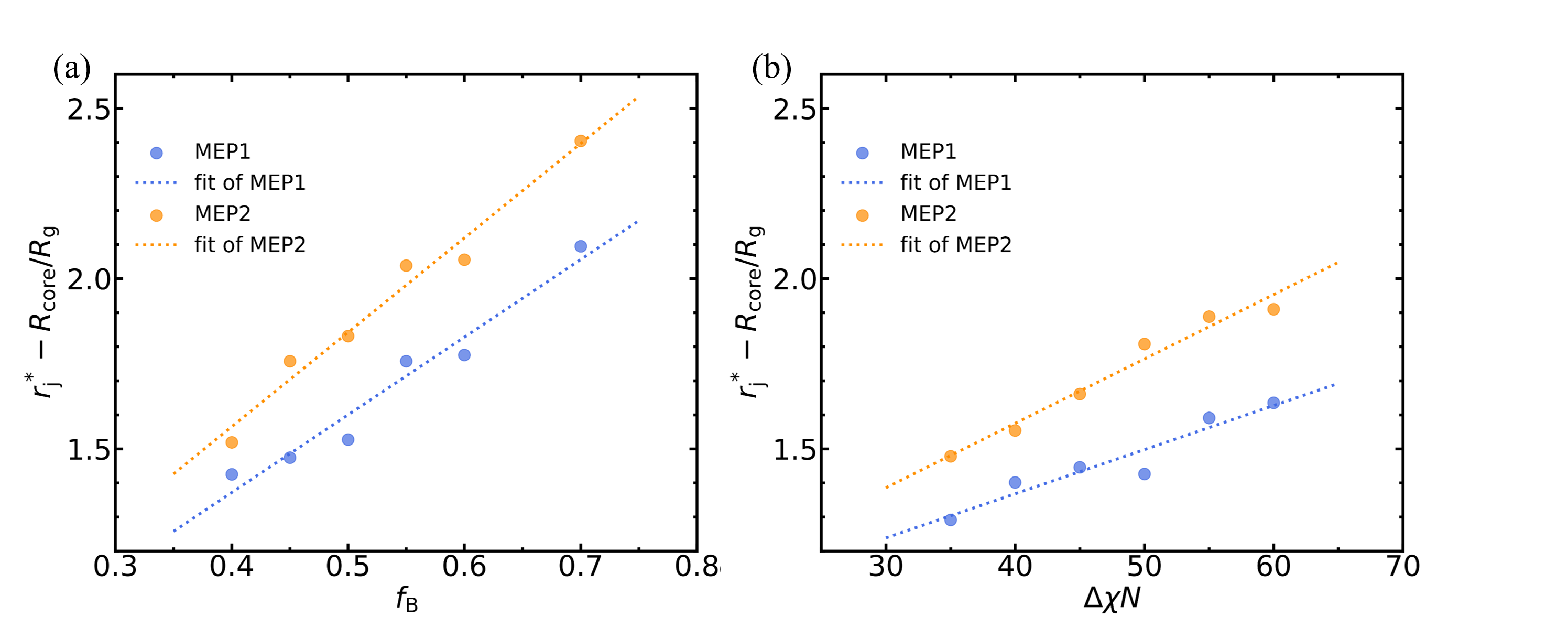}
  \end{subfigure} 
  \caption{
    Effect of hydrophobic block length $f_\mathrm{B}$ and solvent selectivity $\Delta \chi N$ on the distance $r_\mathrm{j}^* - R_\mathrm{core}$ of the junction point from the micelle core interface at the transition state, shown in (a) and (b), respectively.
    In both panels, blue and orange points represent $r_\mathrm{j}^* - R_\mathrm{core}$ of the transition states on MEP1 and MEP2, respectively.
    The blue and orange dashed lines correspond to the linear fits for MEP1 and MEP2.
  }
  \label{fig10}
\end{figure}

We further calculated the distance of the junction point from the micelle core interface at the transition state, $r_\mathrm{j}^* - R_\mathrm{core}$, as shown in Figure \ref{fig10}.  
This quantity characterizes the spatial displacement that the junction point must overcome from the micelle core interface during chain expulsion.
For both MEP1 and MEP2, $r_\mathrm{j}^* - R_\mathrm{core}$ increases linearly with both $f_\mathrm{B}$ and $\Delta \chi N$.  
%%%%%%%%%%%%%%%%
An increase in $\Delta \chi N$ enhances the strength of the unfavorable interactions that the chain must overcome during expulsion, requiring the chain to be pulled farther to escape the attraction of the micelle.  
For chains with larger $f_\mathrm{B}$, in addition to the larger area of unfavorable interactions, there is also a greater loss of conformational entropy upon stretching, further increasing $r_\mathrm{j}^* - R_\mathrm{core}$.
%%%%%%%%%%%%%%%%%%
% The linear dependence on $f_\mathrm{B}$ reveals a stretching-dominated conformation at the transition state, consistent with the larger unfavorable contact area and greater conformational entropy loss for chains with larger $f_\mathrm{B}$.
% The weaker linear dependence on $\Delta \chi N$ reflects the enhanced stretching arising from stronger solvent selectivity, where a larger $\Delta \chi N$ corresponds to stronger unfavorable interactions and hence a larger spatial displacement.
%%%%%%%%%%%%%%%%%%%

\begin{figure}[htbp]
  \centering
  \begin{subfigure}[b]{0.99\textwidth}
    \includegraphics[width=\textwidth]{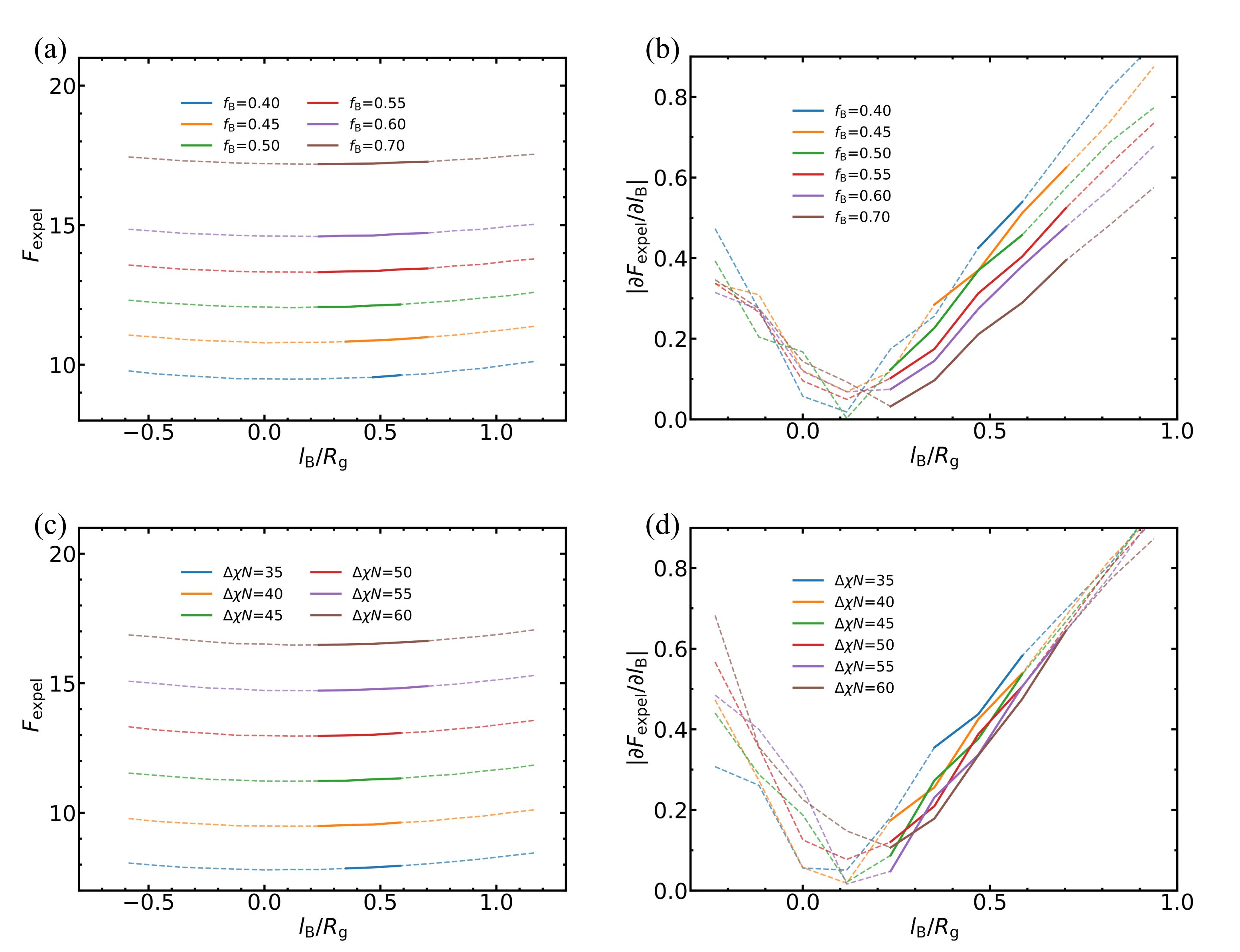}
  \end{subfigure} 
  \caption{
    (a) Free energy profile along the ridge for different hydrophobic block length $f_\mathrm{B}$; 
    (b) corresponding free energy gradient $|\partial F_\mathrm{expel}/ \partial l_\mathrm{B}|$ for different $f_\mathrm{B}$.
    (c) Free energy profile along the ridge for different solvent selectivity $\Delta \chi N$; 
    (d) corresponding free energy gradient for different $\Delta \chi N$.
    The solid lines represent the reaction channel within MEP1 and MEP2, and the dashed line represents the free energy profile along the ridge.  
  }
  \label{fig11}
\end{figure}

To further analyze the reaction channel, we extracted the free energy profile along the ridge on the two-dimensional free energy landscape for the two systems [Figure \ref{fig11}(a),(c)], and marked the transition state region of the reaction channel between MEP1 and MEP2 with solid lines.
The results show that the reaction channel is located in a flat region with the lowest free energy along the ridge, where different conformations are nearly degenerate.  
The system spontaneously selects a set of transition state conformations with low energy barriers.

We further calculated the absolute value of the slope of the ridge free energy profile with respect to the degree of stretching of the hydrophobic block $l_{\mathrm{B}}$, $|\partial F_\mathrm{expel}/ \partial l_\mathrm{B}|$, as shown in Figure \ref{fig11}(b) and (d).  
There is a minimum point, which corresponds to the point of lowest free energy on the ridge.  
The reaction channel is located to the right of this minimum point, and always satisfies $l_{\mathrm{B}} > 0$.  
This indicates that, during the expulsion process, stretched conformations of the hydrophobic block are more favorable in free energy.

In all systems, the left boundary of the reaction channel terminates near the minimum slope point on the ridge.  
It does not extend toward smaller $l_{\mathrm{B}}$, even though the free energy in that region may be slightly lower than that inside the channel.
The further left the boundary lies, the more compact the corresponding conformations of the hydrophobic block become. 
The high sensitivity of the left boundary suggests that compact conformations are highly unfavorable during expulsion.

We note that in systems with shorter hydrophobic blocks, the left boundary of the reaction channel shifts noticeably to the right.  
This behavior likely originates from the corona density.
In our systems, a shorter hydrophobic block corresponds to a smaller micelle core.  
This leads to a higher corona density and a transition state located closer to the micelle interface.
Thus, the hydrophobic block is exposed to a denser corona near the transition state.  
This prevents the hydrophobic block from adopting more compact conformations and shifts the reaction channel toward larger $l_{\mathrm{B}}$.
In contrast, the right boundary corresponds to the maximally stretched conformation of the hydrophobic block, and no significant dependence on corona density is observed.

%==========================
\subsection{Discussions}

Comparing our results with recent molecular dynamics studies by Seeger \emph{et al.} \cite{Seeger2022} and Varner \emph{et al.} \cite{Varner2025} provides further insight into chain expulsion. 
Although those studies and our SCFT approach differ in methodology, all focus on the free energy landscape, the scaling behavior of the expulsion barrier with core block length, and the chain conformation at the transition state.

The MD studies are based on coarse-grained, particle-based models and can explicitly incorporate chain dynamics and fluctuations. 
Seeger \emph{et al.} \cite{Seeger2022} used molecular dynamics with umbrella sampling and identified a hyperstretching mechanism characterized by a linear scaling of the expulsion barrier with core block length ($N_{\mathrm{core}}$).
Near the transition state, the terminal segment of the hydrophobic block tends to remain in the micelle core to avoid unfavorable contacts with the solvent. This forces the hydrophobic block into a highly stretched conformation.
Varner \emph{et al.} \cite{Varner2025} distinguished two scenarios: strong segregation at low density and high-density melt regime.
They found that the former follows the HA collapse mechanism ($N_{\mathrm{core}}^{2/3}$), while the latter exhibits a ``bead-by-bead'' hyperstretching mechanism, with linear scaling ($N_{\mathrm{core}}$). 
They attributed this difference to the influence of the corona structure on the tendency for chain collapse at the transition state under different environments.  
In the melt, the dense corona prevents the hydrophobic block from collapsing, leading to a stretched conformation.

Our work is based on a continuum mean-field theory and employs a Gaussian chain model. 
Although it does not explicitly include fluctuations, it efficiently constructs detailed free energy landscapes to obtain the conformational evolution during chain expulsion. 
It should be noted that, our model cannot reproduce the collapse scaling of a hydrophobic chain in a uniform external field. 
We therefore describe the post-expulsion conformation as a relaxed state relative to the stretched transition state, rather than as a strictly defined collapsed hydrophobic globule. 

Varner \emph{et al.} \cite{Varner2025} also observed nearly degenerate expulsion pathways on their 2D free energy landscapes and noted that chain exchange involves an ensemble of transition states rather than a single well-defined one. 
This observation, together with the continuous reaction channel identified in our work, suggests that chain expulsion from micelles proceeds via multiple pathways. 
The specific mechanism depends strongly on chain conformational fluctuations and the local environment.
%Our two-dimensional analysis further reveals that the characteristics of the reaction channel are primarily governed by the geometric features, rather than by the barrier height. 
%Increases in both the hydrophobic block length and the degree of hydrophobicity shift the entire reaction channel toward larger $r_{\mathrm{j}}$ (away from the micelle interface).  
%The density of the corona mainly affects the boundary of the reaction channel near the micelle core.

%Our SCFT study and the MD studies reach consistent conclusions despite the differences in theoretical models.
%Under high corona density, the transition state for chain expulsion adopts a stretched conformation and its free energy barrier scales linearly with chain length.
%This conclusion provides a unified theoretical picture for explaining the linear scaling observed in experiments.

%%%%%%%%%%%%%%%%%%%%%%%%%%%
\section{Summary}
\label{sec:summary}

This study focuses on the free energy landscape of single chain expulsion in AB diblock copolymer micelles with dense corona. 
We systematically examined the scaling of the expulsion free energy barrier with respect to chain parameters and the characteristics of the transition state.
The expulsion energy barrier scales linearly with both the length of the hydrophobic block ($f_{\mathrm{B}}$) and the solvent selectivity ($\Delta \chi N$). 
Stronger selectivity and larger unfavorable contact areas increase the enthalpy penalty at the core-solvent interface, thereby raising the free energy barrier.
%The barriers show a weaker dependence on micelle aggregation number and the length of the hydrophilic block.  
%This indicates that the chain expulsion is primarily governed by the intrinsic properties of the chain and interfacial interactions.
This work provides a continuum-level theoretical validation of the linear scaling of the expulsion barrier with both $f_{\mathrm{B}}$ and $\Delta \chi N$. 
The results are consistent with recent experimental findings \cite{Choi2010, WangEn2020} and simulations \cite{Seeger2022, Varner2025}.

By introducing two reaction coordinates to describe the spatial position and hydrophobic chain-conformation of the chain, we constructed a two-dimensional free energy landscape and identified the minimum energy paths. 
This approach reveals that the hydrophobic chain adopts a stretched conformation at the transition state. 
We also observed a nearly degenerate reaction channel near the transition state, indicating that chain expulsion can proceed through multiple pathways with comparable barriers.  
The width and boundary of this reaction channel are significantly influenced by the hydrophobic block length and the corona density.  
Increasing the hydrophobic block length shifts the entire channel away from the micelle interface.
Dense corona restricts the inner boundary near the micelle core.
%These microscopic mechanisms provide theoretical support for understanding the dynamic stability of micelle systems.

% Future work should further explore the influence of corona properties on the expulsion transition state by tuning the characteristics of the hydrophilic block and the solvent. 
Finally, we note that recent methodological advances in SCFT have made it possible to describe the coil-to-globule transition within a field-theoretic framework \cite{Xu2021,Liu2024}. 
In future work, combining these methods with the present approach could provide a unified picture covering both the hyperstretching and collapse regimes of chain expulsion.
The SCFT framework established in this study can be extended to more complex architectures, such as multiblock or ionic copolymers. 
Overall, this work provides a general theoretical approach for studying chain exchange dynamics in micellar systems.

\section*{Supplementary Materials}
See the supplementary material for the convergence behavior of the string method and the effect of step size, effect of hydrophilic segment length on the free energy barrier, and effect of micelle aggregation number on the free energy barrier.

%\section{AUTHOR INFORMATION}
%
%\subsection*{Corresponding Authors}
%\textbf{Jiajia Zhou} -- South China Advanced Institute for Soft Matter Science and Technology, School of Emergent Soft Matter, South China University of Technology, Guangzhou 510640, China; Guangdong Provincial Key Laboratory of Functional and Intelligent Hybrid Materials and Devices, South China University of Technology, Guangzhou 510640, China;\\
%\href{https://orcid.org/0000-0002-2258-6757}{orcid.org/0000-0002-2258-6757}; Email: \href{mailto:zhouj2@scut.edu.cn}{zhouj2@scut.edu.cn}
%
%\subsection*{Authors}
%\textbf{Shuang Yuan} -- South China Advanced Institute for Soft Matter Science and Technology, School of Emergent Soft Matter, South China University of Technology, Guangzhou 510640, China; Guangdong Provincial Key Laboratory of Functional and Intelligent Hybrid Materials and Devices, South China University of Technology, Guangzhou 510640, China; \href{https://orcid.org/}{orcid.org/}

%%%%%%%%%%%%%%%%%%%%%%
\begin{acknowledgments}
This research was supported by the Advanced Materials--National Science and Technology Major Project (2025ZD0614503), the National Natural Science Foundation of China (22373036) and R\&D Program of Guangzhou (2024D03J0007). The computation of this work was supported by Scientific Computing Platform of South China University of Technology.
\end{acknowledgments}
%%%%%%%%%%%%%%%%%%%%%%

%\clearpage
\bibliography{barrier}
%%%%%%%%%%%%%%%%%%%%%%

\clearpage
\setcounter{section}{0}
\setcounter{figure}{0}
\renewcommand{\thefigure}{S\arabic{figure}}

\begin{appendix}
%%%%%%%%%%%%%%%%%%%%%%%%%%%%
\section{Convergence Behavior of the String Method with Step Size}
\label{app:string}

Step sizes ranging from $0.001$ to $0.5$ were tested for the reference system ($\mathrm{A}_{0.8}\mathrm{B}_{0.4}$).  
The step size has little effect on the energy barrier (fluctuation $< 0.05\,k_\mathrm{B}T$), but does affect the width of the reaction channel.  
A step size that is too large (e.g., $0.2$) leads to numerical instability and oscillatory convergence.  
A step size that is too small (e.g., $0.005$) results in random drift due to numerical noise, as the free energy surface is extremely flat.  
The channel width increases with step size.  
The free energy surface is nearly flat in the direction perpendicular to the minimum energy path.  
It provides only a weak restoring force. 
A larger step size allows the path to explore a broader region during each iteration.  
An optimal step size range is approximately $0.01-0.1$.  
The step size of $0.02$ was selected for all calculations.

\begin{figure}[htbp]
  \centering
  \begin{subfigure}[b]{0.90\textwidth}
    \includegraphics[width=\textwidth]{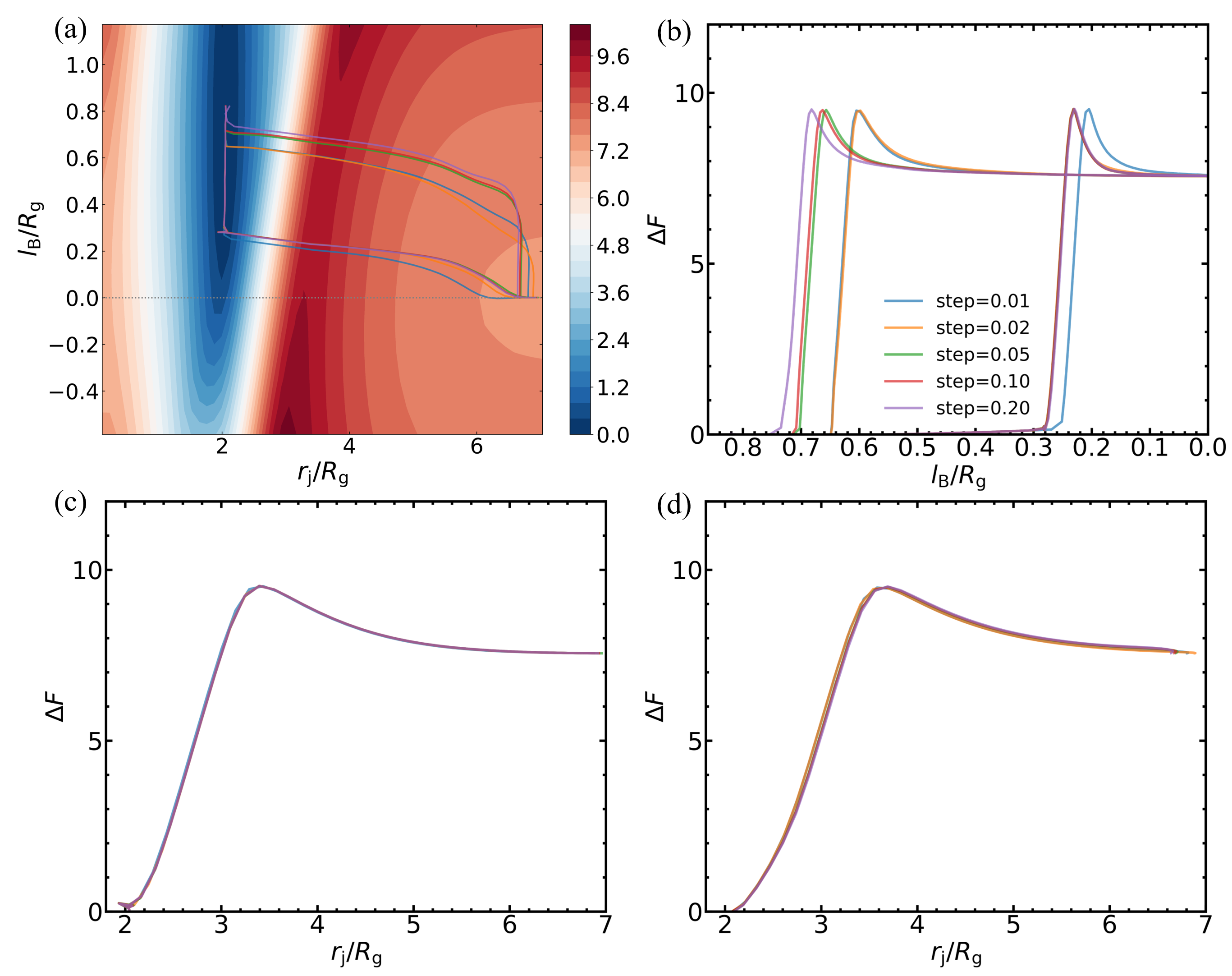}
  \end{subfigure}
  \caption{\linespread{0.9}\selectfont
      Effect of the string method step size on the minimum energy path.
      MEP1 and MEP2 converged with different step sizes, where the line with smaller $l_{\mathrm{B}}$ is MEP1 and the one with larger $l_{\mathrm{B}}$ is MEP2.  
      (a) Two-dimensional free energy surface and converged paths with different step sizes.  
      (b) Free energy as a function of $l_{\mathrm{B}}$ during chain expulsion with different step sizes.  
      (c) and (d) are free energy profiles for MEP1 and MEP2, respectively, with different step sizes. 
    }
  \label{figS1}
\end{figure}

\newpage
%%%%%%%%%%%%%%%%%%%%%%%%%%%%
\section{Effect of Hydrophilic Segment Length}
\label{app:hydrophilic}

We computed the free energy profiles for systems with a fixed hydrophobic block fraction $f_\mathrm{B} = 0.4$ and a variable hydrophilic block fraction $f_\mathrm{A}$ (from $0.4$ to $0.9$). 
The micelle aggregation number was kept as $N_\mathrm{agg}=100$.
The hydrophilic block length has a minor effect on the expulsion barrier ($\sim 0.6\,k_\mathrm{B}T$), but a larger effect on the insertion barrier ($\sim 1.4\,k_\mathrm{B}T$).  
Therefore, in the main text we fix the hydrophilic block length and neglect its effect on the expulsion barrier.

\begin{figure}[htbp]
  \centering
  \begin{subfigure}[b]{0.99\textwidth}
    \includegraphics[width=\textwidth]{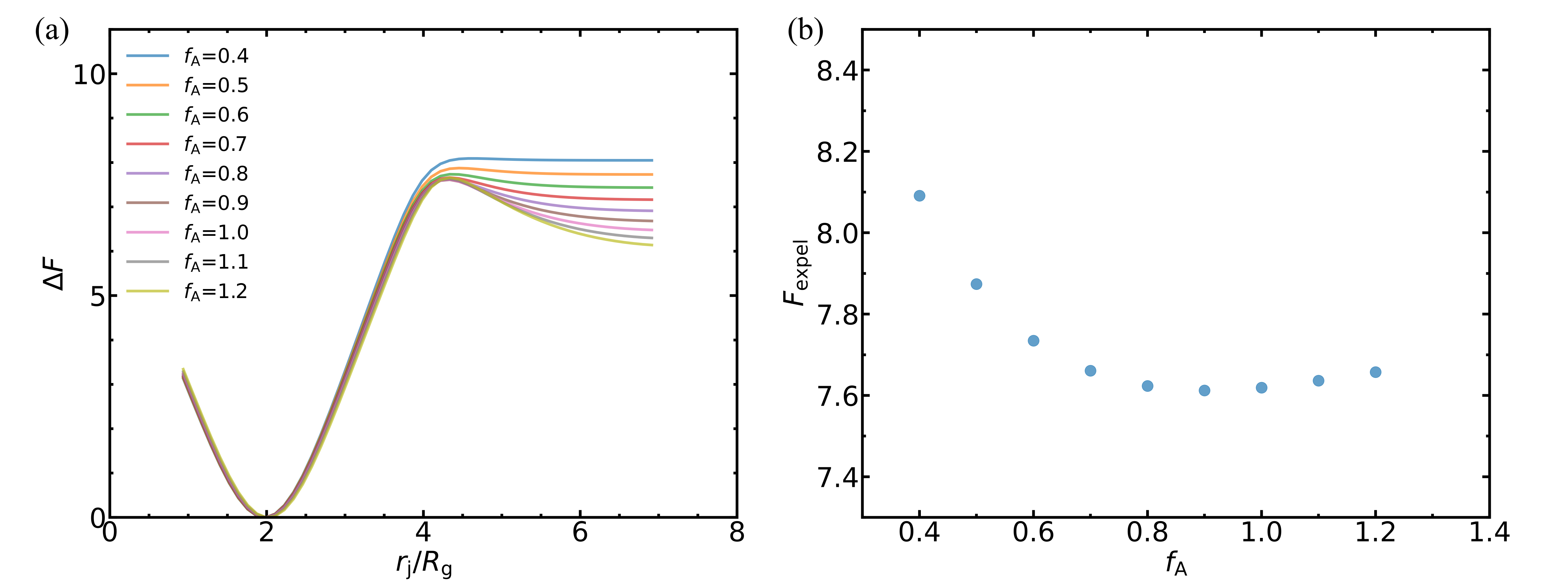}
  \end{subfigure} 
  \caption{\linespread{0.9}\selectfont
    Effect of hydrophilic segment length $f_\mathrm{A}$ on chain exchange free energy.
    (a) Free energy profiles of chain exchange from micelles of variable $f_\mathrm{A}$  (with $f_\mathrm{B}=0.4, N_\mathrm{agg}=100$) as a function of the distance ($r/R_\mathrm{g}$) between the center-of-mass of the micelle and the junction.
    (b) shows the free energy barriers for chain expulsion as a function of $f_\mathrm{A}$. 
    The blue dots represent the free energy barriers.
  }
  \label{figS2}
\end{figure}

\newpage
%%%%%%%%%%%%%%%%%%%%%%%%%%%%
\section{Effect of Micelle Aggregation Number}
\label{app:aggregation_number}

We computed the expulsion free energy profiles for stable micelles with different aggregation numbers. 
For larger micelles, both the reference state and the transition state shift to larger reaction coordinates due to the increased core size.
The shape of the free energy profile remains largely unchanged, and the barrier height increases only weakly with micelle aggregation number. 
Even when the aggregation number doubles, the increase in the free energy barrier is less than $1\,k_\mathrm{B}T$. 
The reduction in micelle aggregation number due to the loss of one expelled chain has a negligible effect on the barrier.

\begin{figure}[htbp]
  \centering
  \begin{subfigure}[b]{0.99\textwidth}
    \includegraphics[width=\textwidth]{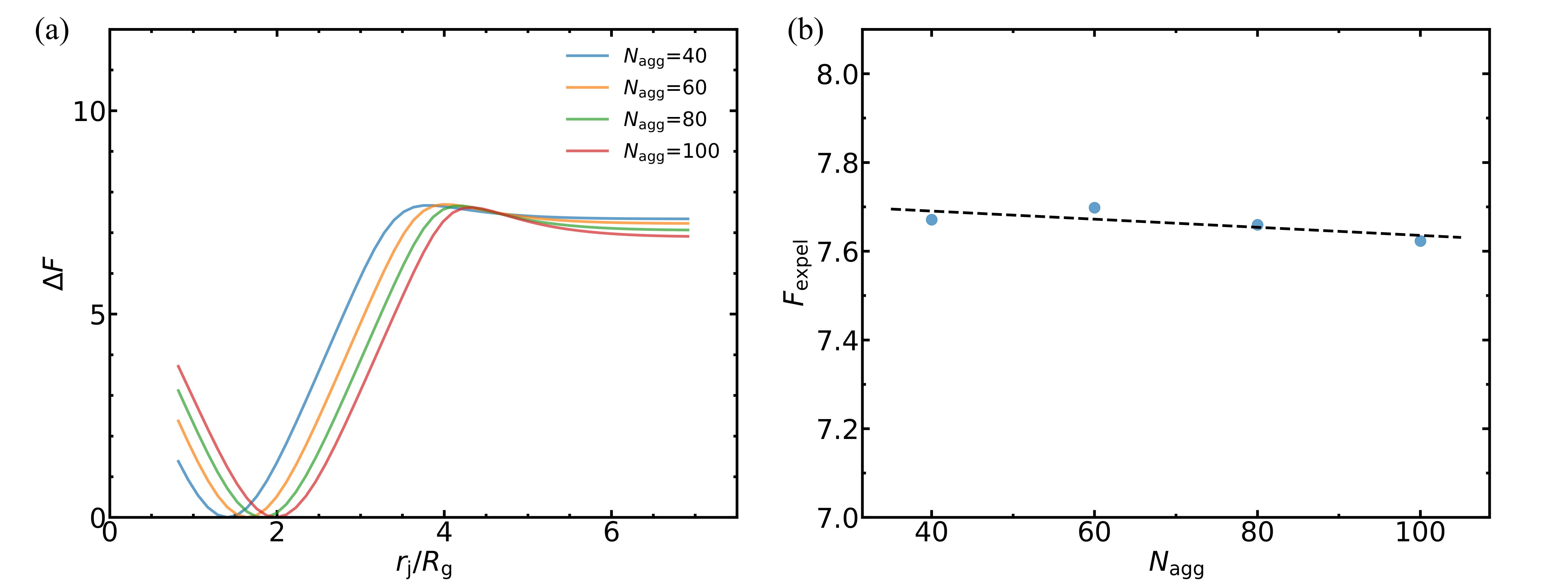}
  \end{subfigure}
  \caption{\linespread{0.9}\selectfont
    Effect of micelle aggregation number $N_\mathrm{agg}$ on the chain exchange free energy.
    (a) Free energy profiles of chain exchange from micelles of different $N_\mathrm{agg}$ (with $f_\mathrm{A}=0.8, f_\mathrm{B}=0.4, \Delta \chi N=40$) as a function of the distance $r/R_\mathrm{g}$ between the micelle center-of-mass and the junction.
    (b) shows the free energy barriers for chain expulsion as a function of $N_\mathrm{agg}$.
    The blue dots represent the free energy barriers, and the dashed line represents the linear fitting.
    }
  \label{figS3}
\end{figure}

\end{appendix}

\end{document}